\documentclass[lettersize,journal]{IEEEtran}
\usepackage{amsmath,amsfonts}
\usepackage{algorithmic}
\usepackage{algorithm}
\usepackage{array}
\usepackage[caption=false,font=normalsize,labelfont=sf,textfont=sf]{subfig}
\usepackage{textcomp}
\usepackage{stfloats}
\usepackage{url}
\usepackage{verbatim}
\usepackage{graphicx}
\usepackage{cite}
\usepackage{multirow}
\usepackage{color}
\usepackage{makecell}
\usepackage{gensymb}
\usepackage[switch]{lineno}
\usepackage{fancyhdr}
\fancypagestyle{firstpage}
{
    \pagestyle{fancy}
    \fancyhead[LO]{\textbf{This work has been submitted to the IEEE for possible publication. Copyright may be transferred without notice, after which this version may no longer be accessible.}}
}

\begin{document}
%\linenumbers

%\title{Data Augmentation With AC-VAEGAN in Sea-Land Clutter Classification for Over-the-Horizon Radar}

\title{Data Augmentation and Classification of
Sea-Land Clutter for Over-the-Horizon Radar Using AC-VAEGAN}

\author{Xiaoxuan Zhang, Zengfu Wang, Kun Lu, and Quan Pan
\thanks{This work was in part supported by the National Natural Science Foundation of China~(grant no. 61790552) and Natural Science Basic Research Plan in Shaanxi Province of China (2021JM-06). }
\thanks{Xiaoxuan Zhang, Zengfu Wang, Quan Pan are with the School of Automation, Northwestern Polytechnical University, and the Key Laboratory of Information Fusion Technology, Ministry of Education, Xi'an, Shaanxi, 710072, China.
Kun Lu is with Nanjing Research Institute of Electronics Technology and the Sky-Rainbow United Laboratory, Nanjing, Jiangsu,
210039, China.
E-mail: (xiaoxuanzhang@mail.nwpu.edu.cn; wangzengfu@nwpu.edu.cn;
mimimomoba@gmail.com;
quanpan@nwpu.edu.cn).
(Corresponding author: Zengfu Wang.)
}
}


\maketitle
\thispagestyle{firstpage}

\begin{abstract}
In the sea-land clutter classification of sky-wave over-the-horizon-radar~(OTHR), the imbalanced and scarce data leads to a poor performance of the deep learning-based classification model.
To solve this problem, this paper proposes an improved auxiliary classifier generative adversarial network~(AC-GAN) architecture,
namely auxiliary classifier variational autoencoder generative adversarial network~(AC-VAEGAN).
AC-VAEGAN can synthesize higher quality sea-land clutter samples than AC-GAN and serve as an effective tool for data augmentation.
Specifically, a 1-dimensional convolutional AC-VAEGAN architecture is designed to synthesize sea-land clutter samples.
\textcolor{black}{Additionally}, an evaluation method combining both traditional evaluation of GAN domain and statistical evaluation of signal domain is proposed to evaluate the quality of synthetic samples.
Using a dataset of OTHR sea-land clutter, both the quality of the synthetic samples and the performance of data augmentation of AC-VAEGAN are verified. Further, the effect of AC-VAEGAN as a data augmentation method on the classification performance of imbalanced and scarce sea-land clutter samples is validated. The experiment results show that the quality of samples synthesized by AC-VAEGAN is better than that of AC-GAN, and the data augmentation method with AC-VAEGAN is able to improve the classification performance in the case of imbalanced and scarce sea-land clutter samples.
\end{abstract}


\section{Introduction}
\label{sec:Introduction}
\IEEEPARstart{A}{s} a crucial system for remote sensing, sky-wave over-the-horizon-radar (OTHR) is widely used in military and civilian fields~\cite{lan2020measurement,yang2020localization,thayaparan2019high,hu2018knowledge}. The sea-land clutter classification of OTHR is the process of identifying whether the background clutter of each range-azimuth cell is originated from land or sea.
Matching the derived classification results with a prior geographic information then provides coordinate registration parameters for target localization, which has the potential to improve target localization accuracy at a low cost~\cite{guo2021improved, yun2003research}.

So far, there have been considerable work on sea-land clutter classification. Turley \emph{et al.}~\cite{turley2013high} defined the energy ratio between the dominant and sub-dominant Bragg resonance peak as a Bragg ratio test statistic, which plays as a key feature for sea-land clutter classification.
Jin \emph{et al.}~\cite{jin2012svm} proposed a support vector machine-based sea-land clutter classification method by analyzing three kinds of features of sea-land clutter.
\textcolor{black}{In ~\cite{turley2013high, jin2012svm},
careful human engineering and considerable domain expertise are needed to design a feature extractor that transform the sea-land clutter spectrum data into a suitable feature vector so that the classifier can classify it into a specific category.
The recently emerging sea-land clutter classification methods based on deep convolutional neural network (DCNN) have achieved remarkable results, of which feature extractor is not designed by human engineers, but learned from data using a general-purpose learning procedure, thus avoiding the drawbacks of the methods in ~\cite{turley2013high, jin2012svm}.}
Li \emph{et al.}~\cite{li2019sea} proposed a DCNN with multiple hidden layers for sea-land clutter classification, of which performance is superior to those of~\cite{turley2013high, jin2012svm}. Besides, sea-land clutter classification based on spectrum data with multi-resolution and multi-scale characteristics was studied.
Zhang \emph{et al.}~\cite{zhang2022bif} leveraged feature maps of different levels to classify sea-land clutter at different scales, and proposed a DCNN method with cross-scale transfer learning.
Li \emph{et al.}~\cite{li2022cross} proposed a cross-scale DCNN sea-land clutter classification method based on the idea of algebraic multi-grid and interpolation-related image downsampling.
However, the DCNN-based methods in~\cite{li2019sea,zhang2022bif,li2022cross} did not consider the problem of imbalanced or scarce sea-land clutter spectrum data.
\textcolor{black}{In the context of sea-land clutter classification, imbalanced sea-land clutter spectrum data means that the number of training samples of different classes is imbalanced; scarce sea-land clutter spectrum data means that only a small number of labeled training samples are available.}

Existing deep learning-based methods require a large number of labeled training data for the classifier to learn a considerable amount of free parameters for accurate prediction.
Although it is easy to collect a large quantity of sea-land clutter data of OTHR, the manual labeling work is quite cumbersome.
The lack of labeled training data for some or all classes often make it difficult for the classifier to perform optimally.
Therefore, it is necessary to improve the performance of the classifier considering imbalanced and scarce sea-land clutter data.

In image classification, one of the most effective ways to deal with scarce samples is data augmentation~\cite{mikolajczyk2018data,shorten2019survey,li2018data}. Traditional methods for image data augmentation perform a certain of geometric affine transformations on training data, such as translation, rotation, mirroring, clipping and scaling, etc., which can improve the performance of classifiers without actually collecting more data.
However, since the sea-land clutter data is radar echo signal, the above-mentioned methods for image data augmentation may destroy the classification-related features of sea-land clutter.
One straightforward way to deal with imbalanced samples is random sampling, which includes random oversampling and random undersampling~\cite{johnson2019survey,buda2018systematic,zhang2021spectral}. Random oversampling refers to randomly copying samples from the minority class to balance original dataset. Since some samples appear repeatedly in the randomly oversampled dataset, the classification model is prone to overfitting. Conversely, random undersampling refers to randomly dropping samples from the majority class to balance original dataset. Since the randomly undersampled dataset may lose some useful information, the classification model only learns part of the overall mode.

As an unsupervised and semi-supervised implicit probability density generative model, generative adversarial network~(GAN) has been successfully applied to data augmentation because of its powerful data synthesis capability~\cite{antoniou2017data,bousmalis2017unsupervised}.
GAN learns data intrinsic distribution through an adversarial game of generator and discriminator to synthesize fake samples that are consistent with original data distribution~\cite{goodfellow2014generative}.
The first fully connected generator and discriminator adversarial architecture provides a theoretical guarantee for the subsequent development of GAN~\cite{goodfellow2014generative}.
Up to now, GAN has evolved into a variety of new architectures.
Deep convolutional generative adversarial network~\cite{radford2015unsupervised} combines DCNN architecture with GAN architecture, which improves the training stability and the quality of samples synthesized by GAN.
Variational autoencoder generative adversarial network~(VAEGAN) \cite{larsen2016autoencoding} is an extension of GAN of which decoder and generator share the same network, while improving the quality of samples synthesized by variational autoencoder (VAE) and GAN.
As unsupervised generative models, however, the GANs in~\cite{goodfellow2014generative, radford2015unsupervised, larsen2016autoencoding}  cannot guide networks to synthesize specific class.
Conditional generative adversarial network~(CGAN)~\cite{mirza2014conditional} performs supervised training by introducing class attribute, which can control the network to synthesize sample with specific class.
Semi-supervised generative adversarial network (SGAN)~\cite{odena2016semi} extends supervised and unsupervised GAN architectures to a semi-supervised architecture, and realizes the synthesis and classification of samples by training a small amount of labeled data and a large amount of unlabeled data.
Auxiliary classifier generative adversarial network~(AC-GAN) \cite{odena2017conditional} combines the network characteristics of CGAN and SGAN, and can output the class which the synthetic sample belongs to while conditionally synthesizing the sample.
On the basis of AC-GAN, improved AC-GAN architectures are developed for data augmentation to solve the problem of imbalanced and scarce samples.
Data augmentation generative adversarial network~(DAGAN)~\cite{antoniou2017data}, based on image CGAN, takes data from a source domain and learns to take any data item and generalize it to generate other within-class data items, which can perform data augmentation on scarce samples.
Balancing generative adversarial network~(BAGAN)~\cite{mariani2018bagan} applies class conditioning in latent space to drive the generation process towards minority class, which can perform data augmentation on imbalanced samples.

The samples synthesized by GAN have potential to balance the minority class of unbalanced and scarce sea-land clutter samples, which can then act as a powerful supplement for the data augmentation of sea-land clutter classification.
To this end, we propose an improved AC-GAN architecture via combining VAEGAN with AC-GAN, namely, auxiliary classifier variational autoencoder generative adversarial network~(AC-VAEGAN).
The overall network structure of AC-VAEGAN is the same as that of VAEGAN, which consists of three parts: encoder~(En), decoder~(De)/generator~(G), and discriminator~(D)/classifier~(C).
The difference is that the input of G of AC-GAN is the combination of random noise sequence and class attribute,
while the proposed AC-VAEGAN attaches an additional input combination of random noise sequence encoded by En of VAEGAN and class attribute.
Since the random noise sequence of the latter is obtained by coding of original data, the combination of the coding sequence and the class attribute is better than that of the former, which makes the training of AC-VAEGAN  more efficient than that of AC-GAN.
On the other hand, AC-VAEGAN can be regarded as an improved VAEGAN, which can specify the class while synthesizing samples.
The proposed AC-VAEGAN is then applied to data augmentation of sea-land clutter classification of OTHR.

The quality of synthetic samples is particularly important.
The sea-land clutter samples are radar echo signals, of which performance cannot be evaluated only by the methods from GAN domain.
To this end, we propose a combination of traditional evaluation of GAN domain and statistical evaluation of signal domain to evaluate the quality of synthetic samples.
From the perspective of evaluation of GAN domain, GAN-train and GAN-test~\cite{shmelkov2018good} are adopted to evaluate the diversity and fidelity of synthetic samples.
From the perspective of statistical evaluation, absolute distance (AD), cosine similarity (CS) and Pearson correlation coefficient (PCC), are proposed to evaluate the quality of synthetic samples from three aspects including signal energy difference, signal direction difference, and signal correlation.

In experiment validation, we prepare three kinds of sea-land clutter datasets: original dataset, imbalanced dataset, and scarce dataset. Based on original dataset, the quality of samples synthesized by AC-VAEGAN and AC-GAN is evaluated comparatively.
Experiment results show that the quality of samples synthesized by AC-VAEGAN is better than that of AC-GAN in terms of the above evaluation metrics.
Furthermore, the effectiveness of AC-VAEGAN data augmentation method is verified based on imbalanced and scarce datasets.
\textcolor{black}{
In summary, our main contributions are four-fold:}
\begin{itemize}
\item \textcolor{black}{For the first time, a novel GAN architecture, namely AC-VAEGAN, is proposed. AC-VAEGAN has the advantages over either of  AC-GAN/VAEGAN architecture, enabling VAEGAN to synthesize sample with specified class, and AC-GAN to synthesize samples with higher quality;
this method is of independent interest.
}

\item \textcolor{black}{An evaluation method combining traditional evaluation of GAN domain and statistical evaluation of signal domain is proposed to evaluate the quality of synthetic samples. The evaluation metrics include GAN-train, GAN-test, AD, CS, and PCC.
}

\item \textcolor{black}{The proposed AC-VAEGAN is successfully applied to data augmentation of imbalanced and scarce sea-land clutter samples of OTHR.
To our best knowledge,
the problem of data augmentation of imbalanced and scarce sea-land clutter samples of OTHR has not been considered in the literature.
The classification model trained with the balanced dataset after data augmentation performs significantly better than that trained with the imbalanced and scarce datasets,
leading to potential improvement on
the coordinate registration of OTHR.
}

\item \textcolor{black}{Also, we verify the data augmentation and classification performance of AC-VAEGAN on the publicly available Moving and Stationary Target Acquisition and Recognition~(MSTAR) dataset. This further demonstrates that the general AC-VAEGAN classification method is applicable not only to one-dimensional remote sensing signal dataset~(e.g., sea-land clutter dataset), but also to two-dimensional remote sensing image dataset~(e.g., MSTAR dataset).
}

\end{itemize}

%It should be emphasized that the proposed AC-VAEGAN is not limited to the data augmentation of sea-land clutter samples of OTHR.
%It can also be applied to other remote sensing fields, such as SAR images, by adapting the convolutional kernel.

The remainder of this paper is organized as follows. In section \ref{sec:Sea-Land Clutter Sample Synthesis Using AC-VAEGAN}, the architecture, sub-module details and training procedure of the proposed AC-VAEGAN are described. In Section \ref{sec:Sea-Land Clutter Sample Evaluation Method}, the evaluation method is proposed to evaluate the quality of synthetic samples. In Section \ref{sec:Experiment and Evaluation}, sea-land clutter dataset and MSTAR dataset are introduced, and then the quality of samples synthesized by AC-VAEGAN and AC-GAN is compared.
Also, data augmentation experiment is performed to verify the effectiveness of AC-VAEGAN.
Section \ref{sec:Conclusions and Future work} draws the conclusions and highlight some future work.

\section{Sea-Land Clutter Sample Synthesis Using AC-VAEGAN}
\label{sec:Sea-Land Clutter Sample Synthesis Using AC-VAEGAN}
In this section, we first describe the proposed AC-VAEGAN architecture and the loss function, and then elaborate on the AC-VAEGAN sub-modules (En, De/G and D/C) for sea-land clutter sample synthesis. At last, we introduce the training and algorithmic procedures of AC-VAEGAN.

\subsection{AC-VAEGAN Architecture}
\begin{figure}[!t]
\centering
\includegraphics[width=2in]{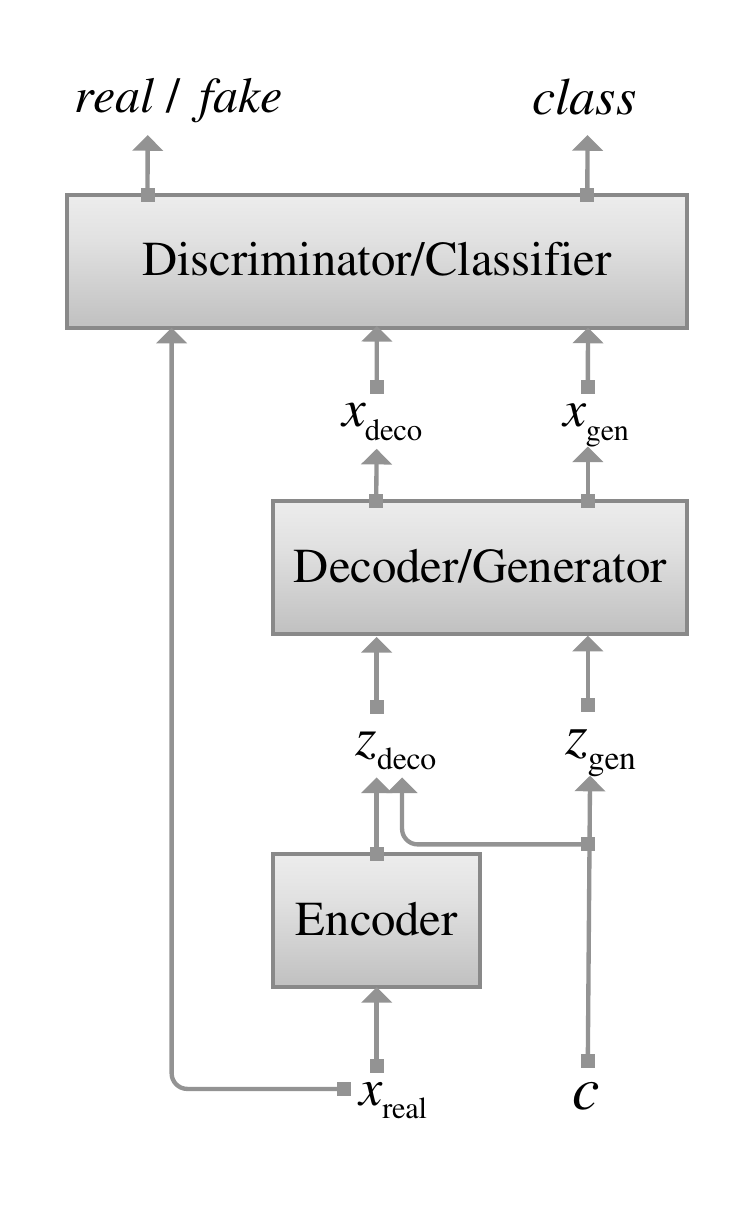}
\caption{The architecture of AC-VAEGAN.}
\label{fig:AC-VAEGAN}
\end{figure}

The architecture of AC-VAEGAN is shown in Fig.~\ref{fig:AC-VAEGAN}.
AC-VAEGAN is a semi-supervised generative model consisting of three parts: En, De/G and D/C.
AC-VAEGAN has the following advantages.
Comparing with VAEGAN, AC-VAEGAN can not only synthesize samples, but also specify the class of synthetic samples. Comparing with AC-GAN, the inputs of G of AC-VAEGAN contain not only the combination of random noise sequence $z_{\text{gen}}$ and class attribute $c$,
but also the combination of $z_{\text{deco}}$ encoded by En of VAEGAN and class attribute $c$, which implies the distribution of $x_{\text{real}}$ with class $c$ in latent space.
Thus, the training of AC-VAEGAN is more efficient than that of AC-GAN.
Also, AC-VAEGAN is easier to synthesize high quality samples than AC-GAN. The loss function of AC-VAEGAN is defined as follows:
\begin{equation}
\label{loss}
L = L_{\text{VAE}} + L_{\text{De/G}} + L_{\text{D/C}},
\end{equation}
where,
\begin{align*}
& L_{\text{VAE}} \!=\! \min \text{MSE}(x_{\text{deco}}, x_{\text{real}})
\!+\! \text{KLD}[N(\mu_{x_{\text{real}}}, \sigma_{x_{\text{real}}}^{2}), N(0, 1)],\\
& L_{\text{De/G}} = \max L_{C} - L_{S},\\
& L_{\text{D/C}} = \max L_{C} + L_{S},\\
& L_{C} = \mathbb{E}[\log p(C = c\vert x_{\text{real}})]
+ \mathbb{E}[\log p(C = c\vert x_{\text{deco}}, x_{\text{gen}})],\\
& L_{S} \!=\! \mathbb{E}[\log p(S = \text{real}\vert x_{\text{real}})]
\!+\! \mathbb{E}[\log p(S = \text{fake}\vert x_{\text{deco}}, x_{\text{gen}})],
\end{align*}
where $x_{\text{real}}$ are real samples, $x_{\text{deco}}$ are fake samples synthesized by De, $x_{\text{gen}}$ are fake samples synthesized by G, $N(0, 1)$ represents the standard normal distribution, $N(\mu_{x_{\text{real}}}, \sigma_{x_{\text{real}}}^{2})$ represents the normal distribution with mean $\mu_{x_{\text{real}}}$ and variance $\sigma_{x_{\text{real}}}^{2}$, $\text{MSE}$ denotes the mean squared error operator, $\text{KLD}$ denotes the Kullback–Leibler divergence operator, $L_{C}$ denotes the log-likelihood of the correct class, and $L_{S}$ denotes the log-likelihood of the correct source.

The input of AC-VAEGAN is 1-dimensional sea-land clutter samples of size $N$, for example, $N=512$.
See Section \ref{subsec:Sea-Land Clutter Dataset} for the details of the sea-land clutter samples.
Since the configuration of sub-modules is related to the size of the input signal,
we next depict the details of the sub-modules by assuming $N=512$.
If $N \neq 512$, the number of layers in the configuration of sub-modules will be adapted to fit the size of the input signal.
The output of AC-VAEGAN include 1-dimensional real or fake probability. and 3-dimensional classification results.

\subsection{Details of AC-VAEGAN Sub-Modules}
Next, we elaborate on AC-VAEGAN sub-modules (En, De/G and D/C) for sea-land clutter sample  synthesis.

\begin{table}[!t]
\centering
\caption{The configuration of AC-VAEGAN sub-modules}
\begin{tabular}{c|c|c|c}
\hline
\hline
Module & Layer & Configuration & Output Size \\
\hline
\multirow{4}*{En} & 1 & Conv1D, LeakyReLU & 8$\times$256 \\
~ & 2-7 & Conv1D, BN1D, LeakyReLU & 512$\times$4 \\
~ & 8 & FC, LeakyReLU & 1000 \\
~ & 9 & FC & 100 \\
\hline
\multirow{4}*{De/G} & 1 & DeConv1D, BN1D, ReLU & 512$\times$4 \\
~ & 2-7 & DeConv1D, BN1D, ReLU & 8$\times$256 \\
~ & 8 & DeConv1D, Tanh & 1$\times$512 \\
\hline
\multirow{4}*{D/C} & 1 & Conv1D, LeakyReLU & 8$\times$256 \\
~ & 2-7 & Conv1D, BN1D, LeakyReLU & 512$\times$4 \\
~ & 8 & FC, LeakyReLU & 1000 \\
~ & 9\_1 & FC, Sigmoid & 1 \\
~ & 9\_2 & FC, Softmax & 3 \\
\hline
\hline
\end{tabular}
\label{table:sub-modules}
\end{table}

\subsubsection{Encoder}
\begin{figure}[!t]
\centering
\includegraphics[width=3in]{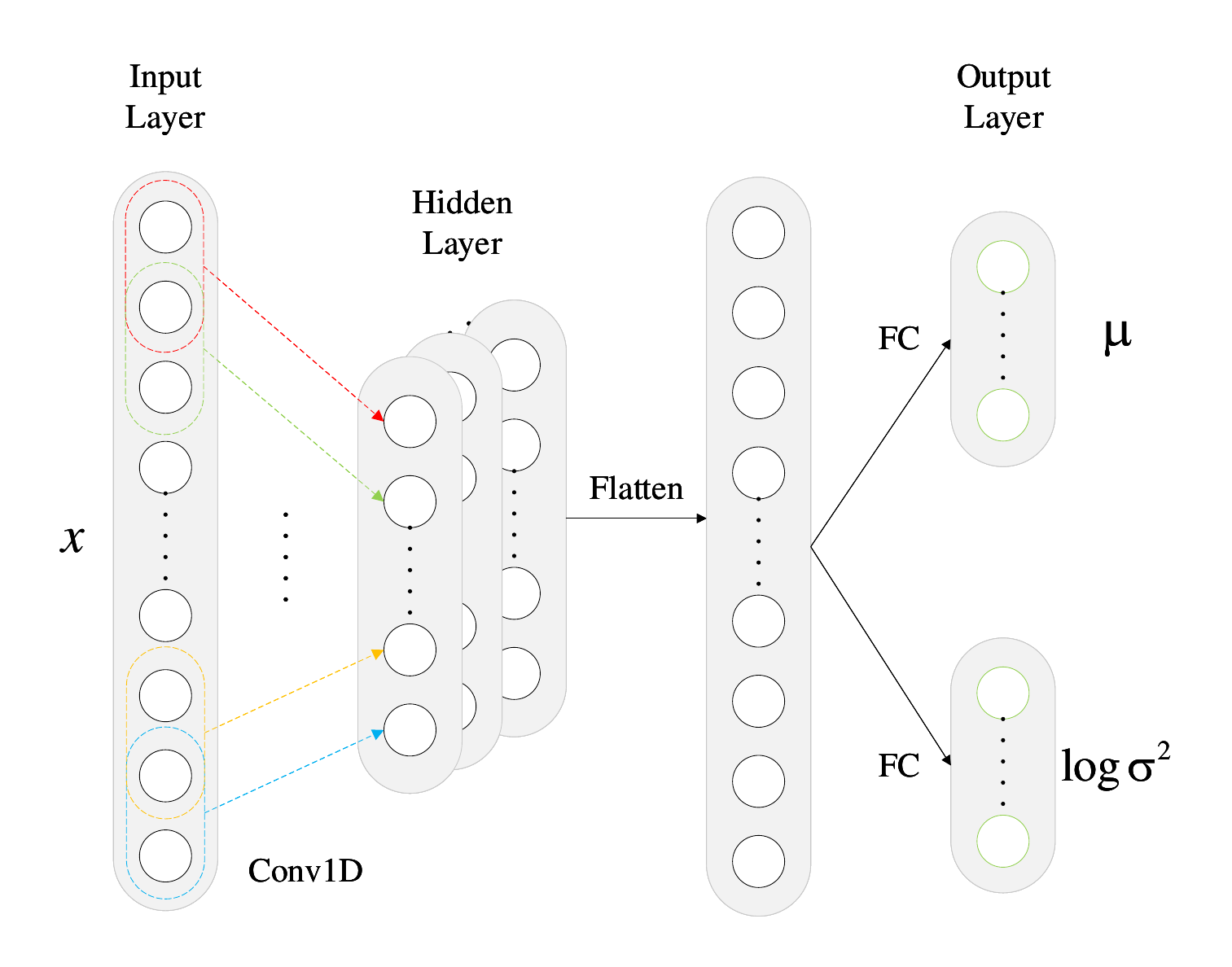}
\caption{Details of En sub-module.}
\label{fig:Encoder}
\end{figure}

En consists of a 9-layer neural network: seven 1-dimensional convolutional (Conv1D) layers are followed by two fully connected (FC) layers.
The architecture and configuration of En are shown in Fig. \ref{fig:Encoder} and Table \ref{table:sub-modules}, respectively.
All convolutional kernels have the same configuration with size 4, stride 2 and padding 1.
As we mentioned above, the input is a real signal of size 1$\times$512.
After layer 1, the number of channels is increased by eight times, and the signal size is decreased by one-half.
Then after layers 2-7, the number of channels is increased by two times per layer, and the signal size is decreased by one-half per layer. Next, flatten the resulting feature vector.
Finally, after layers 8-9, output a 100-dimensional feature vector, which is used to fit the mean $\mu$ or log variance $\log \sigma^{2}$ of normal distribution.
One can sample a random noise sequence $z$ from $N(0, 1)$.
Leveraging the reparameterization technique~\cite{kingma2014auto},
a 100-dimensional random noise sequence $z_{\text{deco}}$ following $N(\mu, \sigma^{2})$ is obtained.

\subsubsection{Decoder/Generator}
\begin{figure}[!t]
\centering
\includegraphics[width=3in]{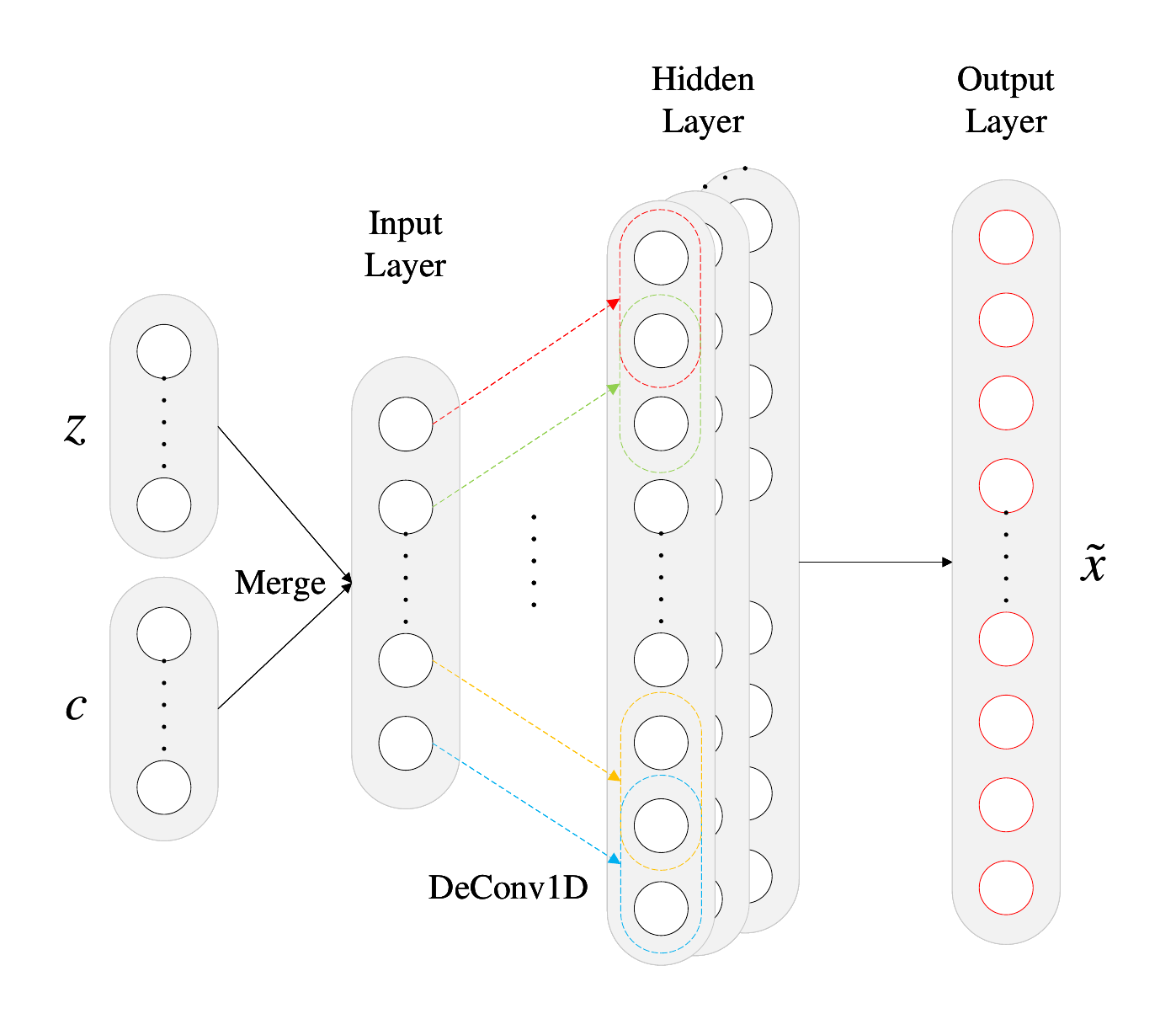}
\caption{Details of De/G sub-module.}
\label{fig:Decoder&Generator}
\end{figure}

De/G consists of a 8-layer neural network: eight 1-dimensional deconvolutional (DeConv1D) layers.
The architecture and configuration of De/G are shown in Fig. \ref{fig:Decoder&Generator} and Table \ref{table:sub-modules}, respectively.
The convolutional kernel of layer 1 has configuration with size 4, stride 1 and padding 0.
The convolutional kernels of layers 2-8 have the same configuration with size 4, stride 2 and padding 1.
\textcolor{black}{The input is a 103-dimensional vector [$z_{\text{deco}}$, $c$] or [$z_{\text{gen}}$, $c$], where $z_{\text{deco}}$ and $z_{\text{gen}}$ are 100-dimensional random noise sequences, and $c$ is a 3-dimensional one-hot encoding of class attribute.}
After layer 1, the number of channels becomes 512, and the dimension of the feature vector is four.
Then after layers 2-7, the number of channels is decreased by one-half per layer, and the signal size is increased by two times per layer.
Finally, after layer 8, output a $1\times512$ reconstructed signal $x_{\text{deco}}$ or generated signal $x_{\text{gen}}$.

\subsubsection{Discriminator/Classifier}
\begin{figure}[!t]
\centering
\includegraphics[width=3in]{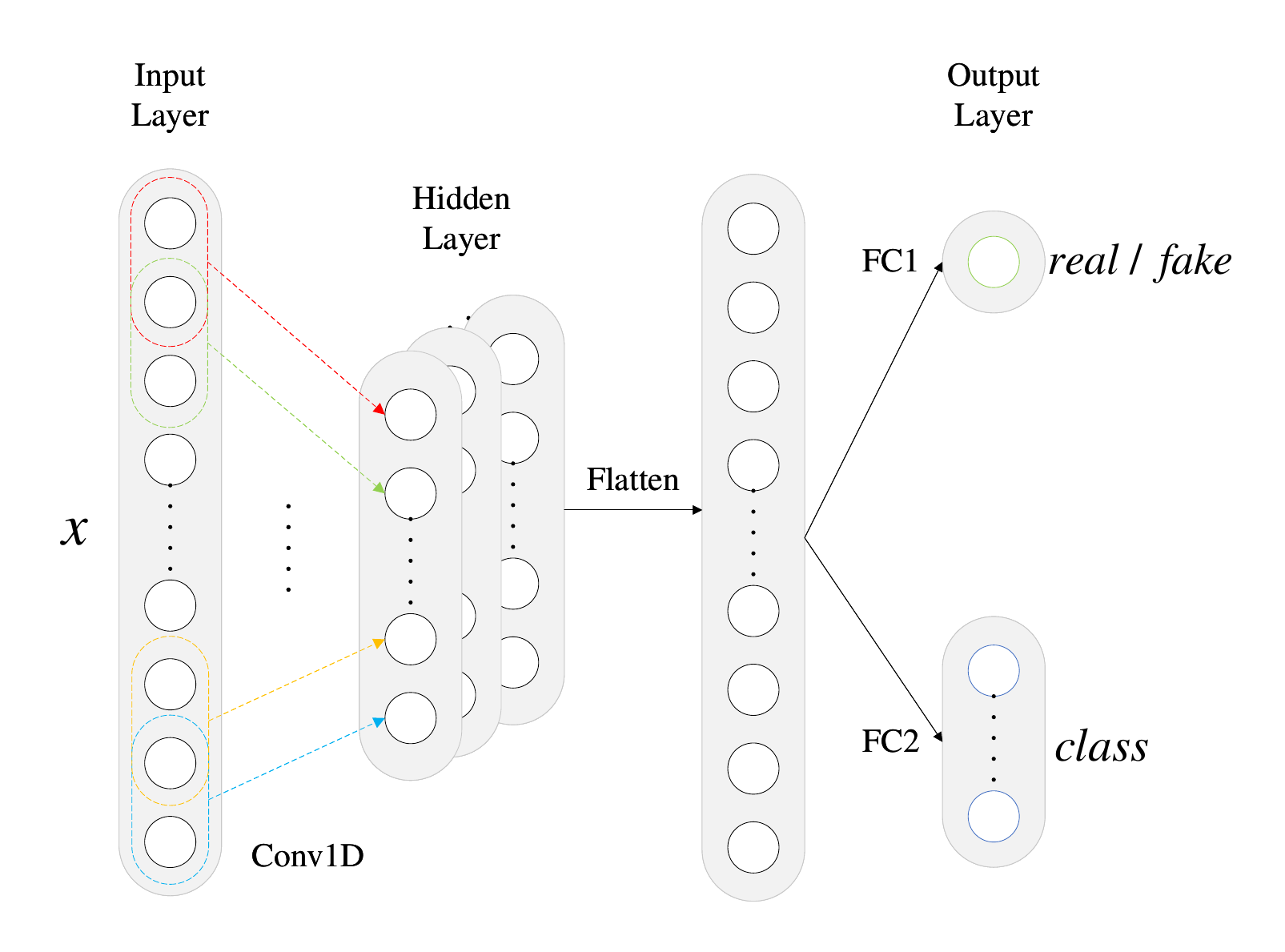}
\caption{Details of D/C sub-module.}
\label{fig:Discriminator&Classifier}
\end{figure}

D/C consists of a 9-layer neural network: seven 1-dimensional convolutional layers are followed by two fully connected layers.
The architecture and configuration of D/C are shown in Fig. \ref{fig:Discriminator&Classifier} and Table \ref{table:sub-modules}, respectively.
Layers 1-8 are consistent with En.
Layer 9 is divided into two channels 9\_1 and 9\_2.
The input is a $1\times512$ real/reconstructed/generated signal.
After layer 9\_1, output a 1-dimensional real or fake probability.
After layer 9\_2, output a 3-dimensional classification result.

\subsection{Training Procedure of AC-VAEGAN}
It is well known that the core idea of VAE is to maximize the variational lower bound of log-likelihood function through En and De~\cite{kingma2014auto}, while the core idea of GAN is to make the network reach Nash equilibrium through ``zero-sum game" of G and D~\cite{goodfellow2014generative}.
AC-VAEGAN combines these two ideas.
During the training of AC-VAEGAN, the role of En is to learn data representation from the real signal $x_{\text{real}}$ and encode it into low-dimensional latent variable $z_{\text{deco}} = \text{En}(x_{\text{real}})$, the role of De/G is to decode/generate $z_{\text{deco}}$ from $N(\mu, \sigma^{2})$ or $z_{\text{gen}}$ from $N(0, 1)$ together with the fake label $c$ into a high-dimensional fake signal, i.e., the reconstructed signal $x_{\text{deco}} = \text{De}(z_{\text{deco}}, c)$ or the generated signal $x_{\text{gen}} = \text{G}(z_{\text{gen}}, c)$, and the role of D/C is to discriminate $x_{\text{real/deco/gen}}$ from the real/reconstructed/generated signal as a real or fake probability $p_{\text{real/deco/gen}} = \text{D}(x_{\text{real/deco/gen}})$ or classify it as a class label $l_{\text{real/deco/gen}} = \text{C}(x_{\text{real/deco/gen}})$.
The detailed training procedure of AC-VAEGAN is described in Algorithm \ref{alg:AC-VAEGAN}.

\begin{algorithm}[!t]
\caption{Training of AC-VAEGAN.}
\label{algorithm}
\begin{algorithmic}
\STATE \textbf{Initialize:} En, De/G, D/C. Let $K$ be the total number of iterations.
\FOR{$k = 1, \ldots, K$}
\STATE \textbf{Train D/C}:
\STATE (1) Sample $m$ real signals $\{x_{\text{real}}^{1}, x_{\text{real}}^{2},\dots,x_{\text{real}}^{m}\}$ from dataset with real labels $\{c_{real}^{1}, c_{real}^{2},\dots,c_{real}^{m}\}$;
\STATE (2) Encode $m$ codings $\{z_{\text{deco}}^{1}, z_{\text{deco}}^{2},\dots,z_{\text{deco}}^{m}\}$ from En:
\STATE \hspace{3cm} $z_{\text{deco}}^{i} = \text{En}(x_{\text{real}}^{i})$
\STATE (3) Sample $m$ fake labels $\{c_{\text{fake}}^{1}, c_{\text{fake}}^{2},\dots,c_{\text{fake}}^{m}\}$ from label space;
\STATE (4) Decode $m$ fake signals $\{x_{\text{deco}}^{1}, x_{\text{deco}}^{2},\dots,x_{\text{deco}}^{m}\}$ from De:
\STATE \hspace{3cm} $x_{\text{deco}}^{i} = \text{De}(z_{\text{deco}}^{i}, c_{\text{fake}}^{i})$
\STATE (5) Sample $m$ latent vectors $\{z_{\text{gen}}^{1}, z_{\text{gen}}^{2},\dots, z_{\text{gen}}^{m}\}$ from noise prior $p(z)$;
\STATE (6) Generate $m$ fake signals $\{x_{\text{gen}}^{1}, x_{\text{gen}}^{2},\dots,x_{\text{gen}}^{m}\}$ from G:
\STATE \hspace{3cm} $x_{\text{gen}}^{i} = \text{G}(z_{\text{gen}}^{i}, c_{\text{fake}}^{i})$
\STATE (7) Update D/C by $L_{D/C}$.
\STATE \textbf{Train De/G and En}:
\STATE (8) Repeat Steps (2)-(6);
\STATE (9) Update De/G and En by $L_{\text{VAE}} + L_{\text{De/G}}$.
\ENDFOR
\end{algorithmic}
\label{alg:AC-VAEGAN}
\end{algorithm}

\section{Evaluation Method of Synthetic Sea-Land Clutter Samples}
\label{sec:Sea-Land Clutter Sample Evaluation Method}
The sea-land clutter samples synthesized by AC-VAEGAN are used for data augmentation, so the quality of the synthetic samples is particularly important.
To our best knowledge, there is no a unified standard for evaluating the quality of samples synthesized by GAN.
In GAN domain, a number of quantitative evaluation metrics have been proposed to evaluate the quality of synthetic samples, including Inception Scores (IS)~\cite{salimans2016improved}, Mode Scores (MS)~\cite{che2019mode}, Fr\'echet Inception Distance (FID)~\cite{heusel2017gans}, Maximum Mean Discrepancy (MMD) \cite{dziugaite2015training}, Earth Mover’s Distance (EMD) \cite{arjovsky2017wasserstein, gulrajani2017improved}, and GAN-train/GAN-test~\cite{shmelkov2018good}, etc.
Different metrics measure different aspects of GAN.
Abundant empirical experiments have shown that
different conclusions on the quality of the synthetic samples may be draw from different evaluation metrics.
Therefore, it is necessary to choose appropriate metrics for real application scenarios.

In our experiment, the sea-land clutter samples used are spectrum data of OTHR. The above metrics of GAN domain evaluate the quality of synthetic samples from
the visual perspective of an image, which is not entirely appropriate to sea-land clutter samples. To this end, a combination of traditional evaluation of GAN domain and statistical evaluation of signal domain is proposed to evaluate the quality of sea-land clutter samples synthesized by AC-VAEGAN.

\textcolor{black}{Considering traditional evaluation of GAN domain, two metrics,
GAN-train and GAN-test,
are used to evaluate the quality of synthetic sea-land clutter samples.
GAN-train treats synthetic samples as training data to train classifier and real samples as test data to obtain classification accuracy, which is used to evaluate the diversity of synthetic samples.
Intuitively, this measures the difference between the learned~(i.e., synthetic clutter) distribution and the target~(i.e., real clutter) distribution.
A good GAN-train performance shows that the synthetic samples are diverse enough.
However, GAN-train also requires a sufficient fidelity, as otherwise the classifier will be impacted by the sample quality.
GAN-test treats real samples as training data to train classifier and synthetic samples as test data to obtain classification accuracy, which is used to evaluate the fidelity of synthetic samples.
A good GAN-test with a high value denoting that the synthetic samples are a realistic approximation of the distribution of real clutter.
}

Considering statistical evaluation of signal domain, three metrics,
AD, CS and PCC,
are proposed to evaluate the quality of synthetic sea-land clutter samples. As a direct distance metric, AD measures the absolute distance between samples, which is used to evaluate the signal energy difference between synthetic samples and real samples.
The value range of AD is $[0, +\infty)$, and the smaller the value, the better the synthetic samples.
CS measures the directional similarity between samples, which is used to evaluate the difference of signal direction between synthetic samples and real samples.
The value range of CS is $[-1, 1]$, and the larger the value, the better the synthetic samples.
The most common metric used to measure the similarity of radar signals is correlation function.
PCC is introduced to evaluate the degree of linear correlation between synthetic samples and real samples. The value range of PCC is $[-1, 1]$, and the larger the value, the better the synthetic samples.

\section{Experiment and Evaluation}
\label{sec:Experiment and Evaluation}
\textcolor{black}{In this section, we verify the effectiveness of the proposed AC-VAEGAN on two datasets, one of which is sea-land clutter dataset and the other is MSTAR dataset.
The purpose of conducting experiment on MSTAR dataset is to demonstrate that AC-VAEGAN has a wider range of applications.
}

\subsection{Dataset}
\subsubsection{Sea-Land Clutter Dataset}
\label{subsec:Sea-Land Clutter Dataset}
\begin{figure}[!t]
\centering
\includegraphics[width=2.5in]{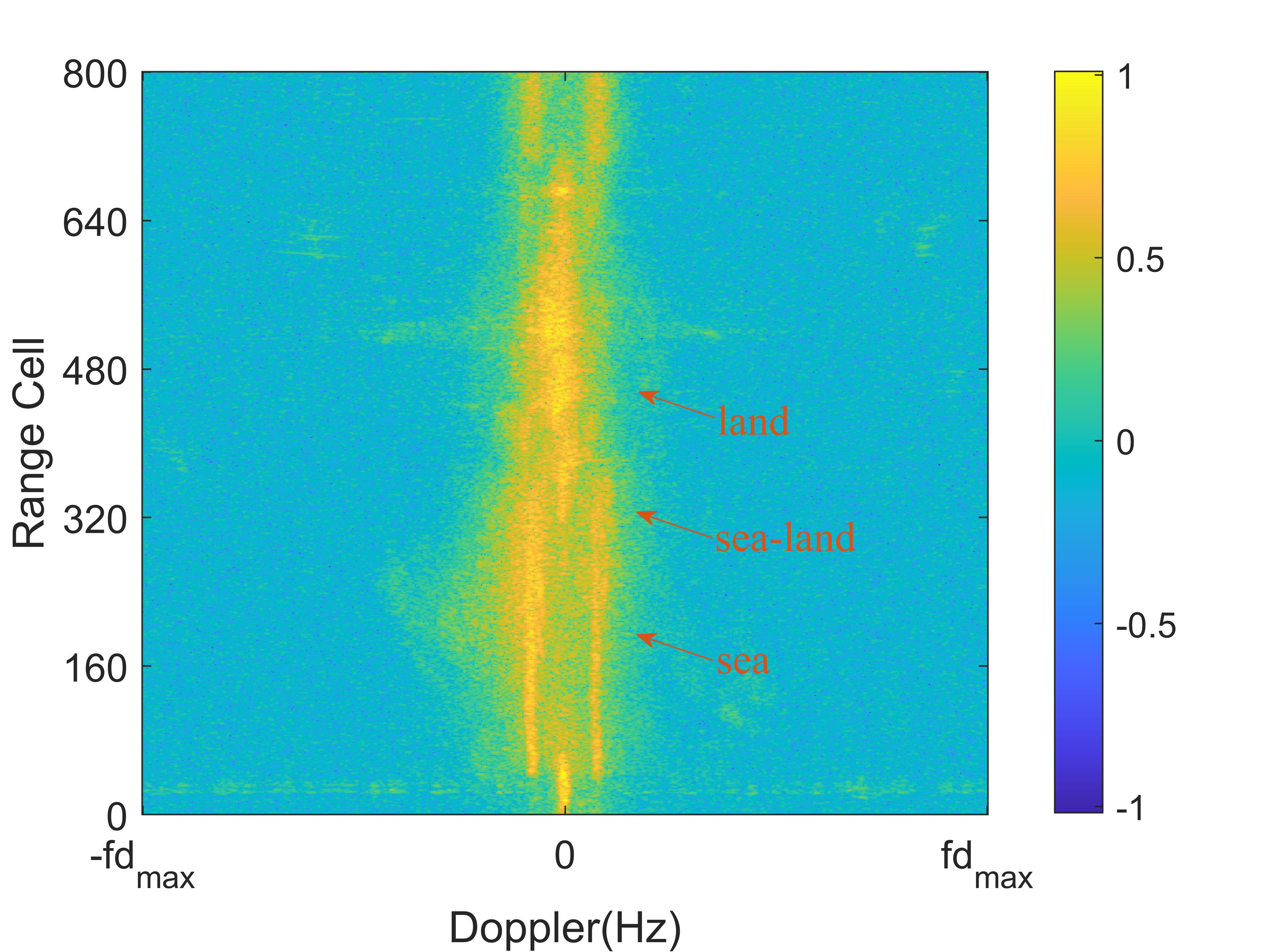}
\caption{The spectrum map of full range of a beam.}
\label{fig:balance spectrum}
\end{figure}

\begin{figure*}[!t]
\centering
\subfloat[]{\includegraphics[width=2in]{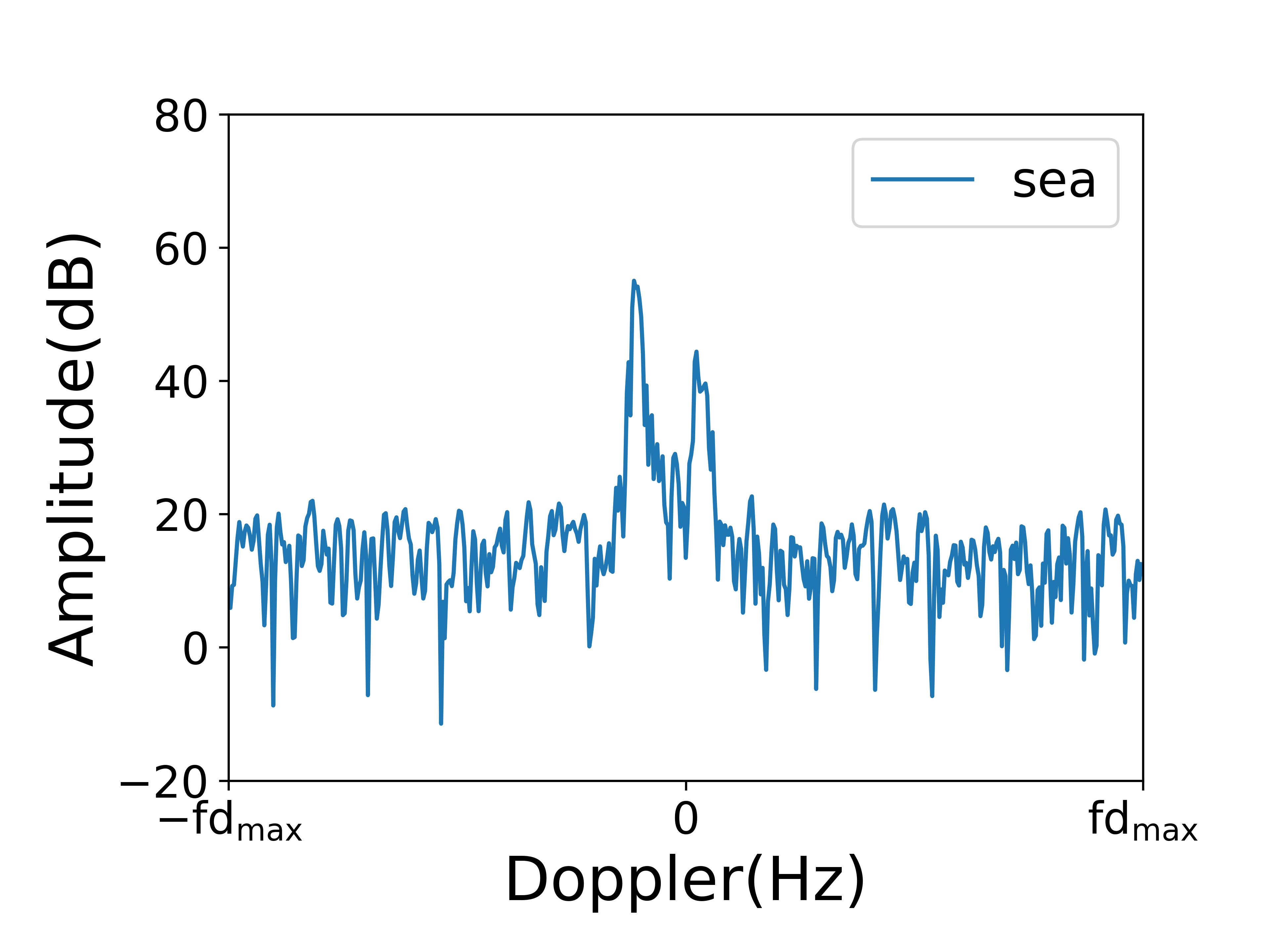}
\label{fig:sea_real}}
\hfil
\subfloat[]{\includegraphics[width=2in]{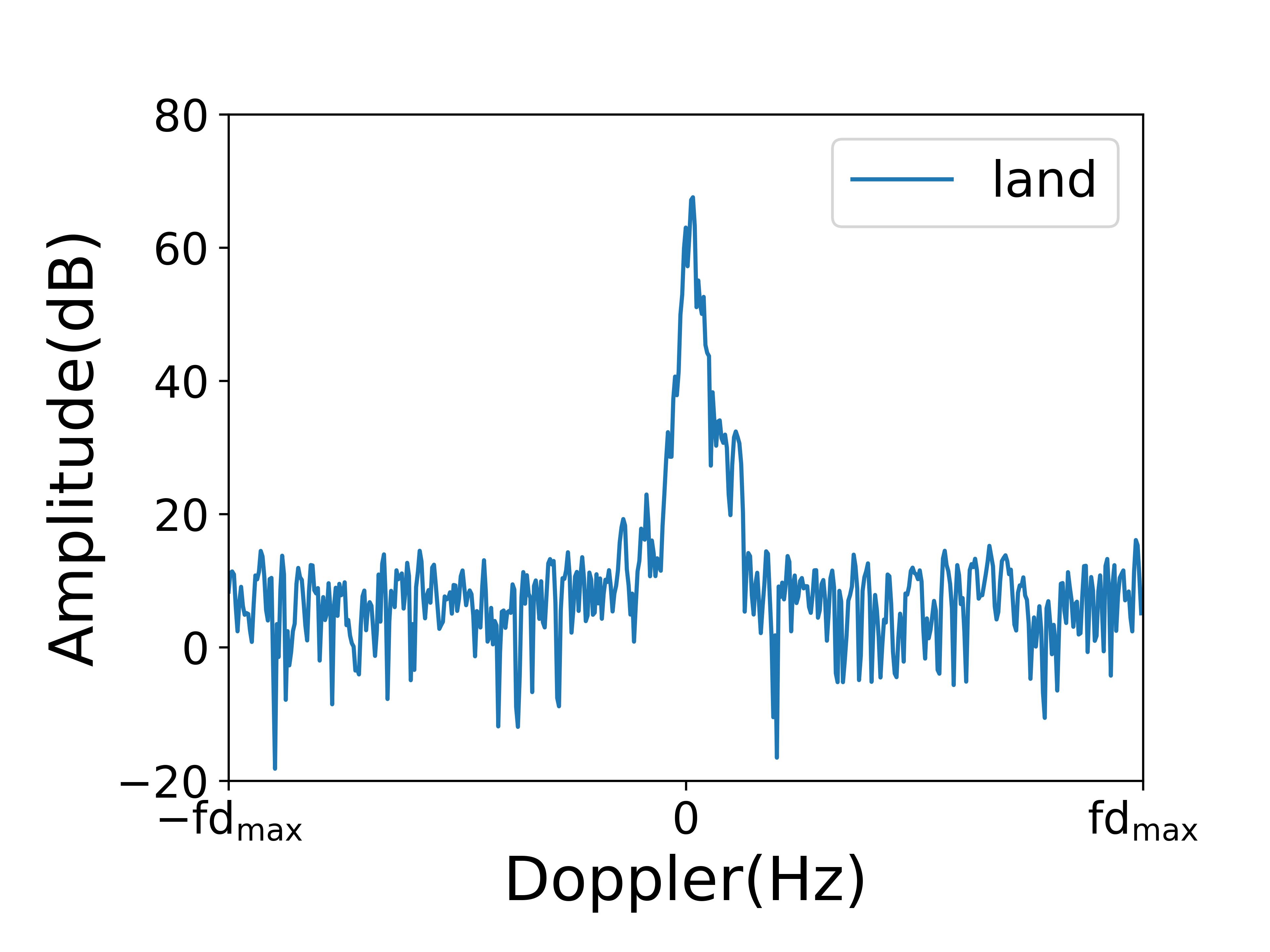}
\label{fig:land_real}}
\hfil
\subfloat[]{\includegraphics[width=2in]{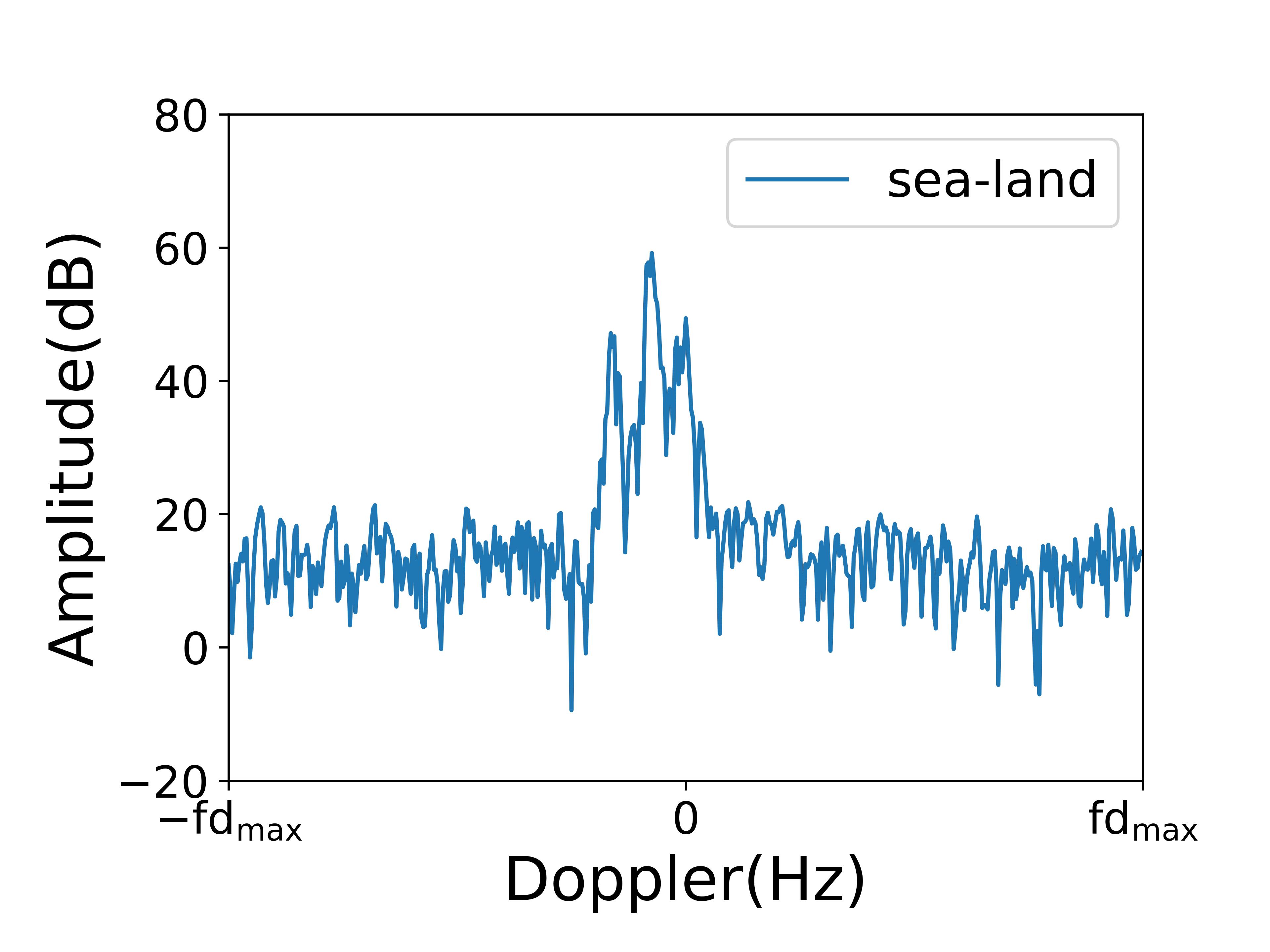}
\label{fig:sea-land_real}}
\caption{Sea-land clutter dataset. (a) An example of sea clutter. (b) An example of land clutter. (c) An example of sea-land boundary clutter.}
\label{fig:real sample}
\end{figure*}

To verify the effectiveness of AC-VAEGAN, we adopt sea-land clutter dataset as the benchmark dataset, which is the spectrum of clutter obtained by OTHR. %Specifically, the electromagnetic wave signals are emitted by the transmitter of OTHR, and are reflected by ionosphere to the surface of sea or land, and then the echo signals of clutter return to the receiver~\cite{headrick1974over}.
\textcolor{black}{
Fig.~\ref{fig:balance spectrum} shows the spectrum map of full range of a beam from OTHR under a good ionospheric condition.
}
We take a collection of spectrum of a range-azimuth cell as sea-land clutter dataset, as shown in Fig.~\ref{fig:real sample}.
Fig.~\ref{fig:real sample}\subref{fig:sea_real} is an example of sea clutter.
The first-order Bragg peak of sea clutter is formed by the Bragg resonance scattering of high-frequency electromagnetic waves emitted by OTHR and ocean waves, showing double peaks symmetrical to zero frequency.
Fig.~\ref{fig:real sample}\subref{fig:land_real} is an example of land clutter.
Since land is stationary, it is observed that a single peak appears near zero frequency.
Fig.~\ref{fig:real sample}\subref{fig:sea-land_real} is an example of sea-land boundary clutter, which combines the features of sea clutter and land clutter, showing triple peaks.
\textcolor{black}{In addition to the examples shown in Fig.~\ref{fig:real sample}, the sea/land clutter dataset also includes
samples under a relatively poor ionospheric condition or complex environment, such as samples with Doppler shifts, spread Doppler spectrum  and radio-frequency interference~(see Fig.~\ref{fig:gen sample}).
Note that there are more complex sea/land clutter if the ionospheric condition is very poor.
In this paper, we limit ourselves to the sea/land clutter obtained under not very poor
ionospheric condition,
that is, the sea/land clutter that is not hard for human experts to identify.
We leave the classification of more complicated sea/land clutter as our future work.}
The detailed description of sea-land clutter dataset used to evaluate AC-VAEGAN is shown in Table \ref{table:balance sample}, in which training data is used for the training of AC-VAEGAN and the evaluation of synthetic samples, and test data is only used for the evaluation of synthetic samples and do not participate in any training process.

\begin{table*}[!t]
\centering
\caption{Detailed Description of Sea-Land Clutter Dataset}
\label{table:balance sample}
\begin{tabular}{cccccc}
\hline
Class & Attribute & Label & Quantity & Training Quantity & Test Quantity\\
\hline
1 & sea & 0 & 1000 & 700(70\%) & 300(30\%)\\
2 & land & 1 & 1000 & 700(70\%) & 300(30\%)\\
3 & sea-land & 2 & 1000 & 700(70\%) & 300(30\%)\\
\hline
\end{tabular}
\end{table*}

Further, we randomly divide the original sea-land clutter dataset into imbalanced and scarce datasets,
in which the test data is consistent with the original test data,
and only some samples of one or more classes are excluded from the original training data.
See Table~\ref{table:imbalance and scarce samples}.
The data augmentation performance of AC-VAEGAN is verified based on the above datasets.

\begin{table}[!t]
\centering
\caption{Sea-Land Clutter Training Data for Data Augmentation}
\label{table:imbalance and scarce samples}
\begin{tabular}{cccc}
\hline
NO. & Sea & Land & Sea-Land\\
\hline
1 & 700(100\%) & 700(100\%) & 140(20\%)\\
2 & 700(100\%) & 140(20\%) & 700(100\%)\\
3 & 140(20\%) & 700(100\%) & 700(100\%)\\
4 & 350(50\%) & 350(50\%) & 350(50\%)\\
5 & 210(30\%) & 210(30\%) & 210(30\%)\\
6 & 140(20\%) & 140(20\%) & 140(20\%)\\
\hline
\end{tabular}
\end{table}

\subsubsection{\textcolor{black}{MSTAR Dataset}} 
\begin{figure*}
\centering
\subfloat[]{
\begin{minipage}[t]{0.15\textwidth}
\centering
\includegraphics[height=0.8in, width=0.8in]{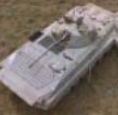}\\
\vspace{0.02cm}
\includegraphics[height=0.8in, width=0.8in]{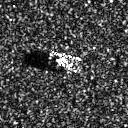}
\end{minipage}}
\subfloat[]{
\begin{minipage}[t]{0.15\textwidth}
\centering
\includegraphics[height=0.8in, width=0.8in]{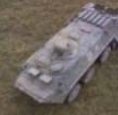}\\
\vspace{0.02cm}
\includegraphics[height=0.8in, width=0.8in]{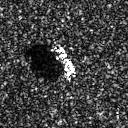}
\end{minipage}}
\subfloat[]{
\begin{minipage}[t]{0.15\textwidth}
\centering
\includegraphics[height=0.8in, width=0.8in]{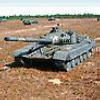}\\
\vspace{0.02cm}
\includegraphics[height=0.8in, width=0.8in]{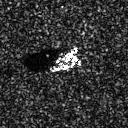}
\end{minipage}}
\subfloat[]{
\begin{minipage}[t]{0.15\textwidth}
\centering
\includegraphics[height=0.8in, width=0.8in]{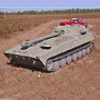}\\
\vspace{0.02cm}
\includegraphics[height=0.8in, width=0.8in]{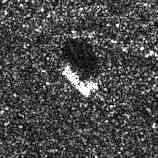}
\end{minipage}}
\subfloat[]{
\begin{minipage}[t]{0.15\textwidth}
\centering
\includegraphics[height=0.8in, width=0.8in]{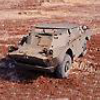}\\
\vspace{0.02cm}
\includegraphics[height=0.8in, width=0.8in]{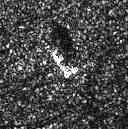}
\end{minipage}}

\vfil
\vspace{1pt}

\subfloat[]{
\begin{minipage}[t]{0.15\textwidth}
\centering
\includegraphics[height=0.8in, width=0.8in]{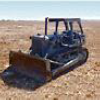}\\
\vspace{0.02cm}
\includegraphics[height=0.8in, width=0.8in]{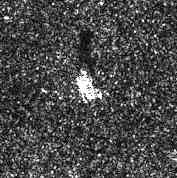}
\end{minipage}}
\subfloat[]{
\begin{minipage}[t]{0.15\textwidth}
\centering
\includegraphics[height=0.8in, width=0.8in]{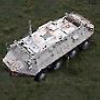}\\
\vspace{0.02cm}
\includegraphics[height=0.8in, width=0.8in]{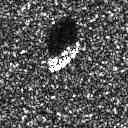}
\end{minipage}}
\subfloat[]{
\begin{minipage}[t]{0.15\textwidth}
\centering
\includegraphics[height=0.8in, width=0.8in]{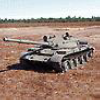}\\
\vspace{0.02cm}
\includegraphics[height=0.8in, width=0.8in]{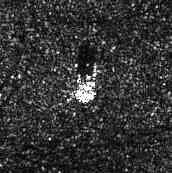}
\end{minipage}}
\subfloat[]{
\begin{minipage}[t]{0.15\textwidth}
\centering
\includegraphics[height=0.8in, width=0.8in]{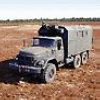}\\
\vspace{0.02cm}
\includegraphics[height=0.8in, width=0.8in]{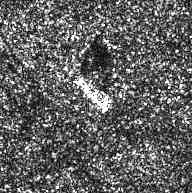}
\end{minipage}}
\subfloat[]{
\begin{minipage}[t]{0.15\textwidth}
\centering
\includegraphics[height=0.8in, width=0.8in]{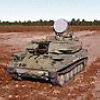}\\
\vspace{0.02cm}
\includegraphics[height=0.8in, width=0.8in]{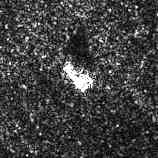}
\end{minipage}}

\caption{\textcolor{black}{MSTAR dataset: optical images~(top) and the corresponding SAR images~(bottom). (a) BMP2. (b) BTR70. (c) T72. (d) 2S1. (e) BRDM2. (f) D7. (g) BTR60. (h) T62. (i) ZIL131. (j) ZSU234.}}
\label{fig:real sample-2}
\end{figure*}

\textcolor{black}{Besides, the effectiveness of AC-VAEGAN is verified on the MSTAR dataset, which was collected by the Sandia National Laboratory~(SNL) in a project jointly sponsored by the Defense Advanced Research Projects Agency~(DARPA) and the Air Force Research Laboratory~(AFRL)~\cite{chen2016target, keydel1996mstar}.
As shown in Fig.~\ref{fig:real sample-2}, the publicly available MSTAR dataset~(128$\times$128 pixels) includes ten different types of ground targets~(BMP2, BTR70, T72, 2S1, BRDM2, D7, BTR60, T62, ZIL131, ZSU234).
It was collected by using an X-band SAR sensor in one foot resolution spotlight mode, full aspect coverage~(in the range from $0\degree$ to $360\degree$).
The detailed description of MSTAR dataset used to evaluate AC-VAEGAN is shown in Table \ref{table:balance sample-2}, in which the data collected at
$17\degree$ depression angles are used for training, and the data collected at $15\degree$ depression angles are used for testing.
}

\textcolor{black}{Further, we randomly divide the original MSTAR dataset into imbalanced and scarce datasets,
in which the test data is consistent with the original test data,
and only some samples of one or more classes are excluded from the original training data.
See Table~\ref{table:imbalance and scarce samples-2}.
The data augmentation performance of AC-VAEGAN is verified based on the above datasets.}

\begin{table*}[!t]
\centering
\caption{\textcolor{black}{Detailed Description of MSTAR Dataset}}
\label{table:balance sample-2}
\begin{tabular}{cccccc}
\hline
Class & Attribute & Label & Quantity & Training Quantity & Test Quantity\\
\hline
1 & BMP2 & 0 & 428 & 233 & 195\\
2 & BTR70 & 1 & 429 & 233 & 196\\
3 & T72 & 2 & 428 & 232 & 196\\
4 & 2S1 & 3 & 573 & 299 & 274\\
5 & BRDM2 & 4 & 572 & 298 & 274\\
6 & D7 & 5 & 573 & 299 & 274\\
7 & BTR60 & 6 & 451 & 256 & 195\\
8 & T62 & 7 & 572 & 299 & 273\\
9 & ZIL131 & 8 & 573 & 299 & 274\\
10 & ZSU234 & 9 & 573 & 299 & 274\\
\hline
\end{tabular}
\end{table*}

\begin{table*}[!t]
\centering
\caption{\textcolor{black}{MSTAR Training Data for Data Augmentation}}
\label{table:imbalance and scarce samples-2}
\begin{tabular}{ccccccccccc}
\hline
NO. & BMP2 & BTR70 & T72 & 2S1 & BRDM2 & D7 & BTR60 & T62 & ZIL131 & ZSU234\\
\hline
1 & 50 & 75 & 100 & 125 & 150 & 175 & 200 & 225 & 250 & 275\\
2 & 50 & 50 & 100 & 100 & 150 & 150 & 200 & 200 & 250 & 250\\
3 & 50 & 50 & 50 & 50 & 50 & 75 & 100 & 125 & 150 & 175\\
4 & 150 & 150 & 150 & 150 & 150 & 150 & 150 & 150 & 150 & 150\\
5 & 100 & 100 & 100 & 100 & 100 & 100 & 100 & 100 & 100 & 100\\
6 & 50 & 50 & 50 & 50 & 50 & 50 & 50 & 50 & 50 & 50\\
\hline
\end{tabular}
\end{table*}

\subsection{Samples Synthesis}
\subsubsection{Synthesis of Sea-Land Clutter Samples}
The experiment environment used in this work for sea-land clutter sample synthesis and subsequent data augmentation is shown in Table \ref{table:enviroment}.

\begin{table}[!t]
\centering
\caption{Experiment Environment}
\label{table:enviroment}
\begin{tabular}{cc}
\hline
Environment & Version\\
\hline
System & Windows 10 (64-bit)\\
GPU & NVIDIA GeForce RTX 3090\\
CUDA & 11.6\\
python & 3.9.0 (in Anaconda 4.11.0)\\
torch & 1.11.0\\
torchvision & 0.12.0\\
numpy & 1.22.3\\
matplotlib & 3.5.1\\
\hline
\end{tabular}
\end{table}

\begin{figure*}[!t]
\centering
\includegraphics[width=6in]{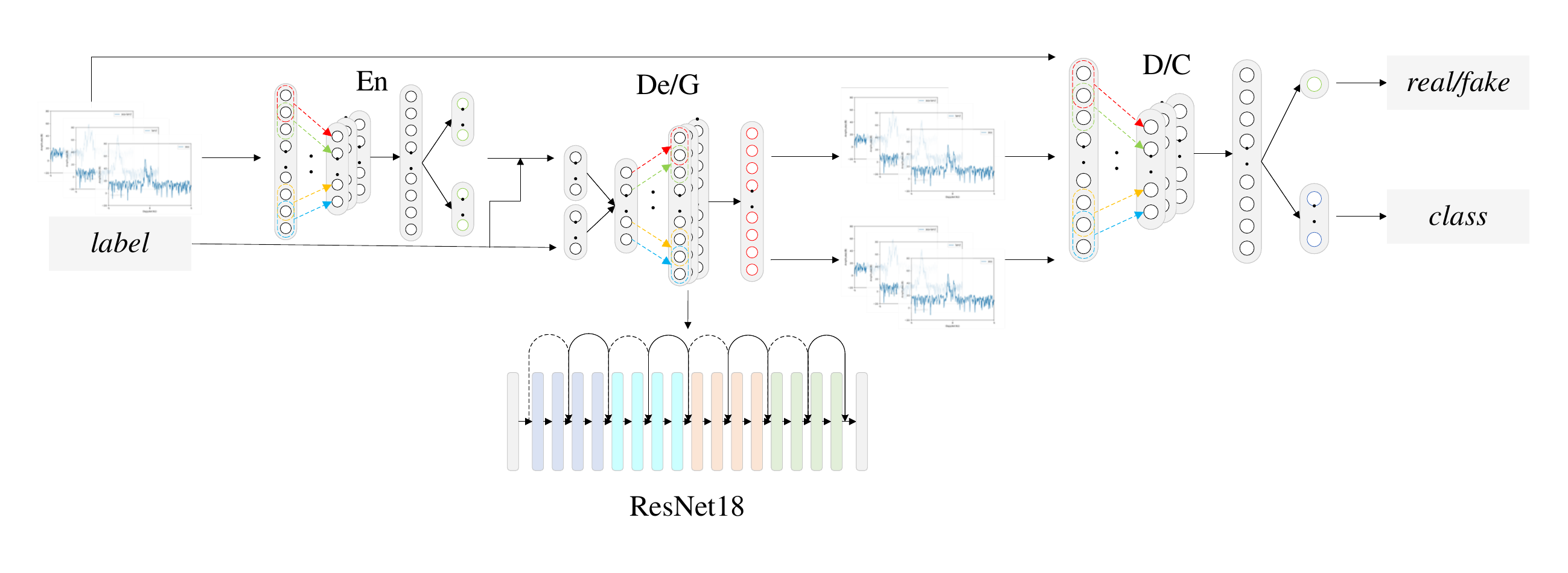}
\caption{The schematic diagram of sea-land clutter sample synthesis and data augmentation.}
\label{fig:schematic diagram}
\end{figure*}

The schematic diagram of sea-land clutter sample synthesis and data augmentation based on AC-VAEGAN is shown in Fig.~\ref{fig:schematic diagram}. The three types of clutter samples in the original sea-land clutter training data of sea, land and sea-land boundary are used as the input of AC-VAEGAN.
By Algorithm~\ref{alg:AC-VAEGAN}, fake clutter samples are synthesized, which are used for subsequent samples quality evaluation and data augmentation.
The hyperparameter configuration of AC-VAEGAN for sea-land clutter sample synthesis is shown in Table \ref{table:paramater synthesis}.

\begin{table}[!t]
\centering
\caption{The hyperparameter Configuration of AC-VAEGAN for Sea-Land Clutter Sample Synthesis}
\label{table:paramater synthesis}
\begin{tabular}{cc}
\hline
Configuration & Default\\
\hline
Training Epoch & 1000\\
Batch Size & 64\\
Learning Rate & 0.0001\\
KLD & -\\
MSE Loss & -\\
BCE Loss & -\\
Cross Entropy Loss & -\\
Adma Optimizer & beta1:0.5, beta2:0.999\\
Data Normalization & [-1,1]\\
Weight Initialization & -\\
\hline
\end{tabular}
\end{table}

\begin{figure*}[!t]
\centering
\subfloat[]{\includegraphics[width=2.5in]{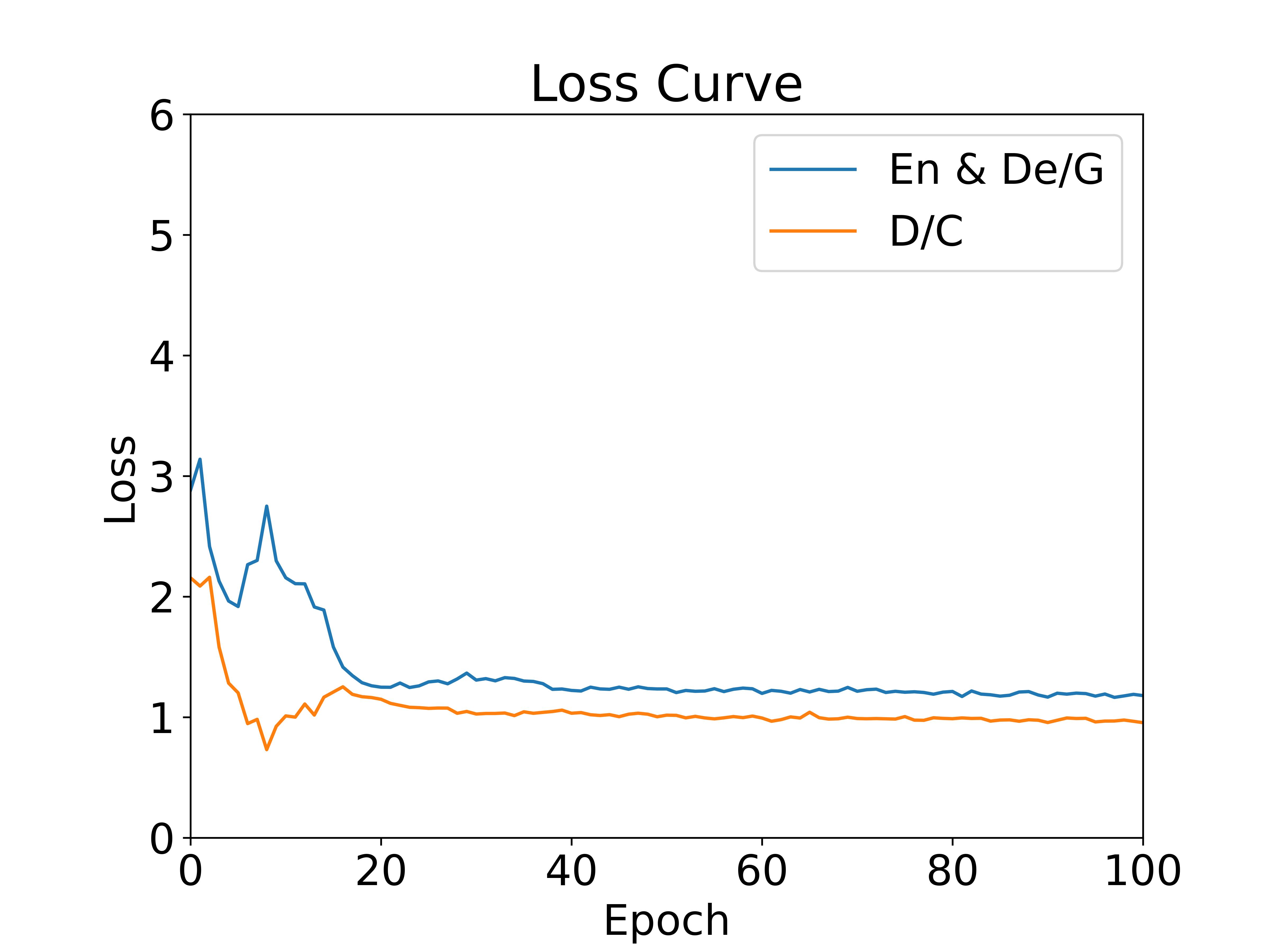}%
\label{fig:loss}}
\hfil
\subfloat[]{\includegraphics[width=2.5in]{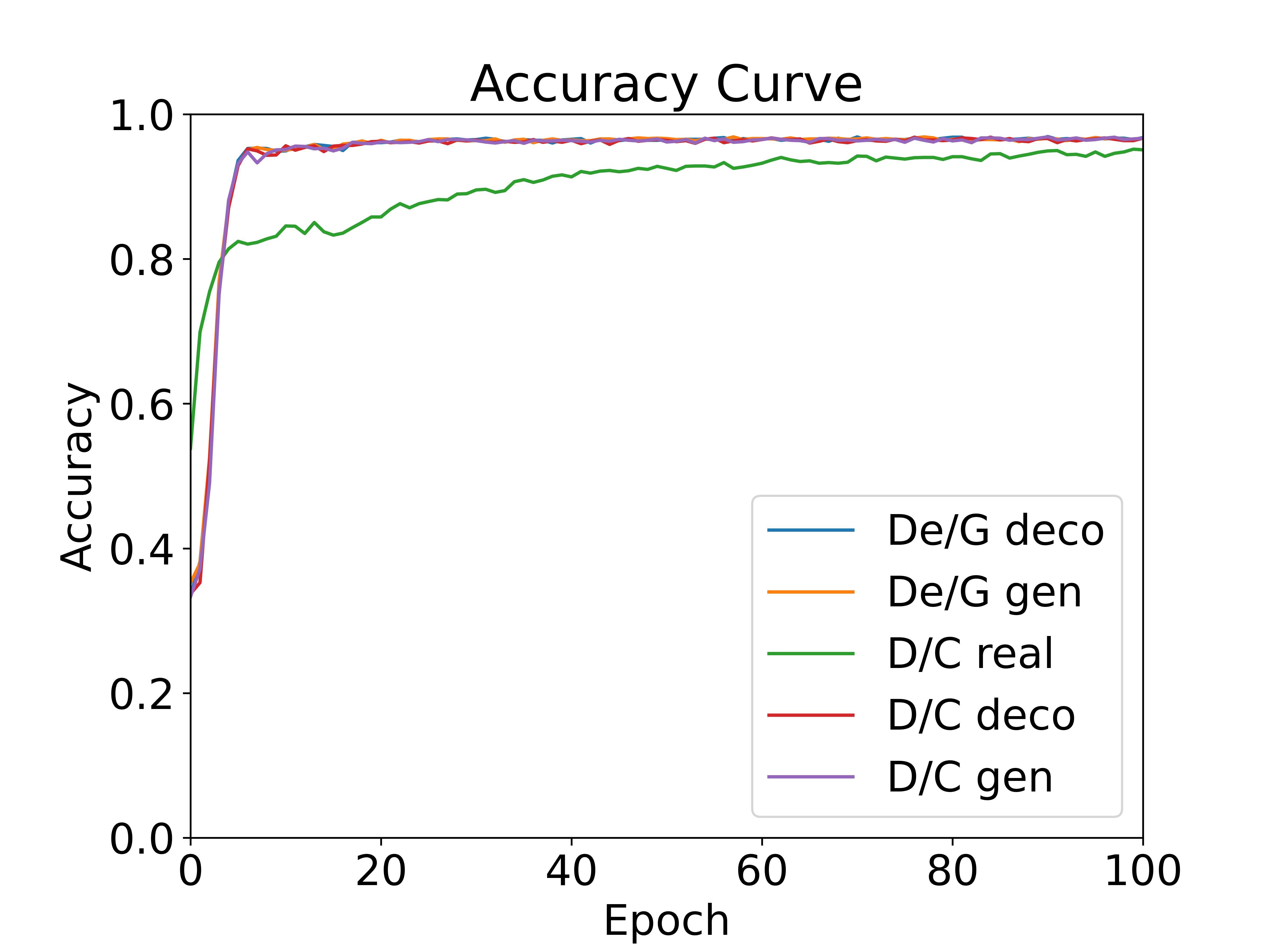}%
\label{fig:accuracy}}
\caption{\textcolor{black}{Loss curve and classification accuracy curve of AC-VAEGAN for sea-land clutter dataset. (a) Loss curve. (b) Accuracy curve.}}
\label{fig:loss-accuracy}
\end{figure*}

\textcolor{black}{
Fig.~\ref{fig:loss-accuracy}\subref{fig:loss} and Fig.~\ref{fig:loss-accuracy}\subref{fig:accuracy} plot the loss curve and the classification accuracy curve of AC-VAEGAN in the first 100 training epochs for sea-land clutter dataset, respectively,
from which we can conclude that the training of AC-VAEGAN is stable and converges rapidly.}

\subsubsection{\textcolor{black}{Synthesis of MSTAR Samples}}
\textcolor{black}{
Firstly, the convolutional kernel of AC-VAEGAN is adjusted to fit the MSTAR samples of size $N=128\times128$.
See Table \ref{table:sub-modules-2} for the corresponding configuration, where the size of all convolutional kernels is 3 in En and D/C, the size of convolutional kernels of layers 1-4 is 5, and the size of convolutional kernel of layer 5 is 8 in De/G.
In addition, we add dropout with the probability of 0.5 in D/C to prevent overfitting.
The ten types of MSTAR samples in the original MSTAR training data of BMP2, BTR70, T72, 2S1, BRDM2, D7, BTR60, T62, ZIL131 and ZSU234 are used as the input of AC-VAEGAN.
By Algorithm~\ref{alg:AC-VAEGAN}, fake clutter samples are synthesized, which are used for subsequent samples quality evaluation and data augmentation.
The hyperparameter configurations of MSTAR samples are the same as those of sea-land clutter samples~(see Table \ref{table:paramater synthesis}).
}

\begin{table}[!t]
\centering
\caption{\textcolor{black}{The configuration of AC-VAEGAN for the MSTAR dataset}}
\begin{tabular}{c|c|c|c}
\hline
\hline
Module & Layer & Configuration & Output Size \\
\hline
\multirow{2}*{En} & 1-6 & Conv2D, ReLU & 512$\times$13$\times$13 \\
~ & 7 & FC & 100 \\
\hline
\multirow{2}*{De/G} & 1-4 & DeConv2D, BN2D, ReLU & 64$\times$61$\times$61 \\
~ & 5 & DeConv2D, Tanh & 1$\times$128$\times$128 \\
\hline
\multirow{4}*{D/C} & 1 & Conv2D, LeakyReLU, Dropout & 16$\times$64$\times$64 \\
~ & 2-6 & \thead{Conv2D, BN2D, LeakyReLU,\\ Dropout} & 512$\times$13$\times$13 \\
~ & 7\_1 & FC, Sigmoid & 1 \\
~ & 7\_2 & FC, Softmax & 10 \\
\hline
\hline
\end{tabular}
\label{table:sub-modules-2}
\end{table}

\begin{figure*}[!t]
\centering
\subfloat[]{\includegraphics[width=2.5in]{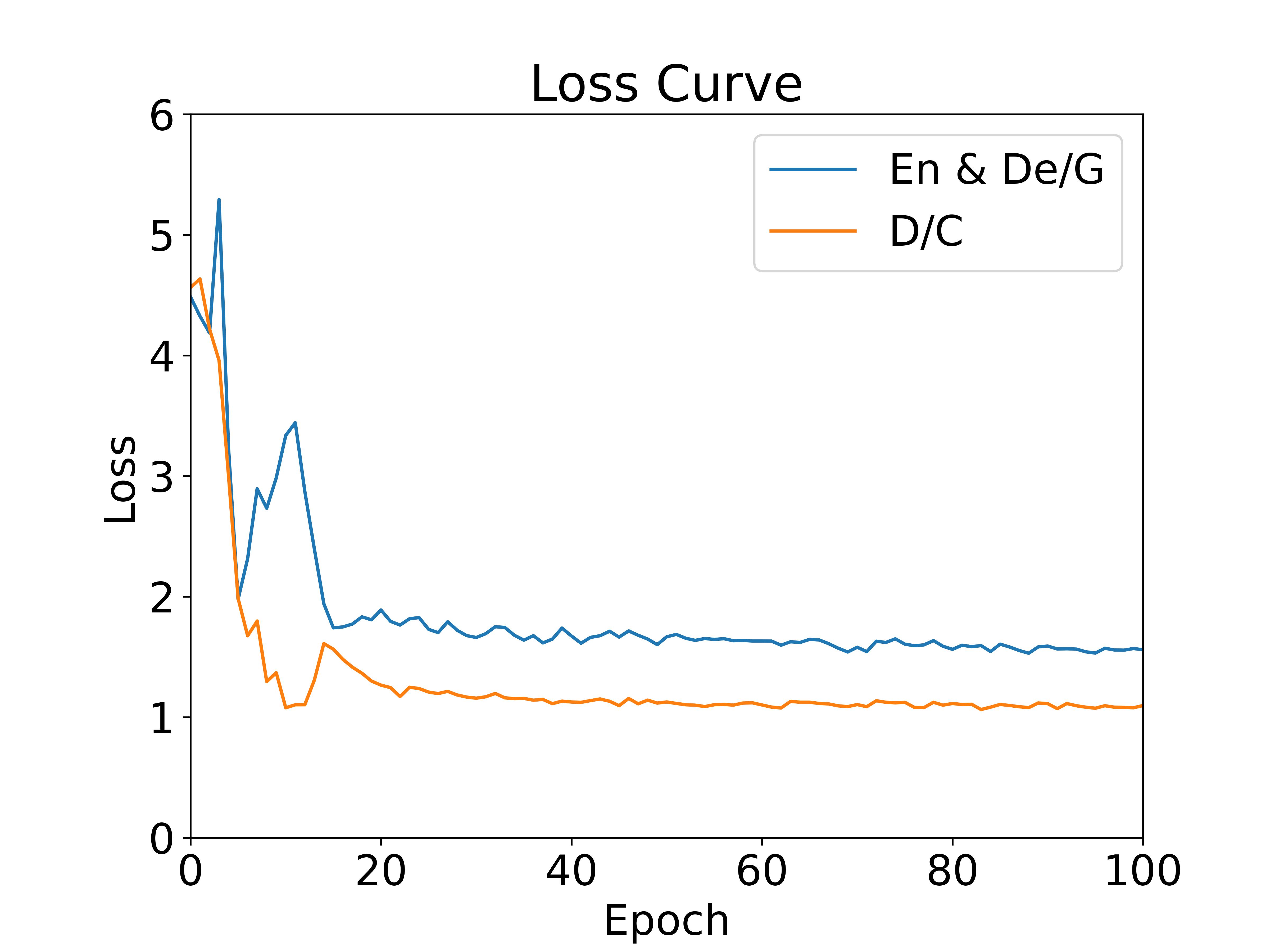}%
\label{fig:loss-2}}
\hfil
\subfloat[]{\includegraphics[width=2.5in]{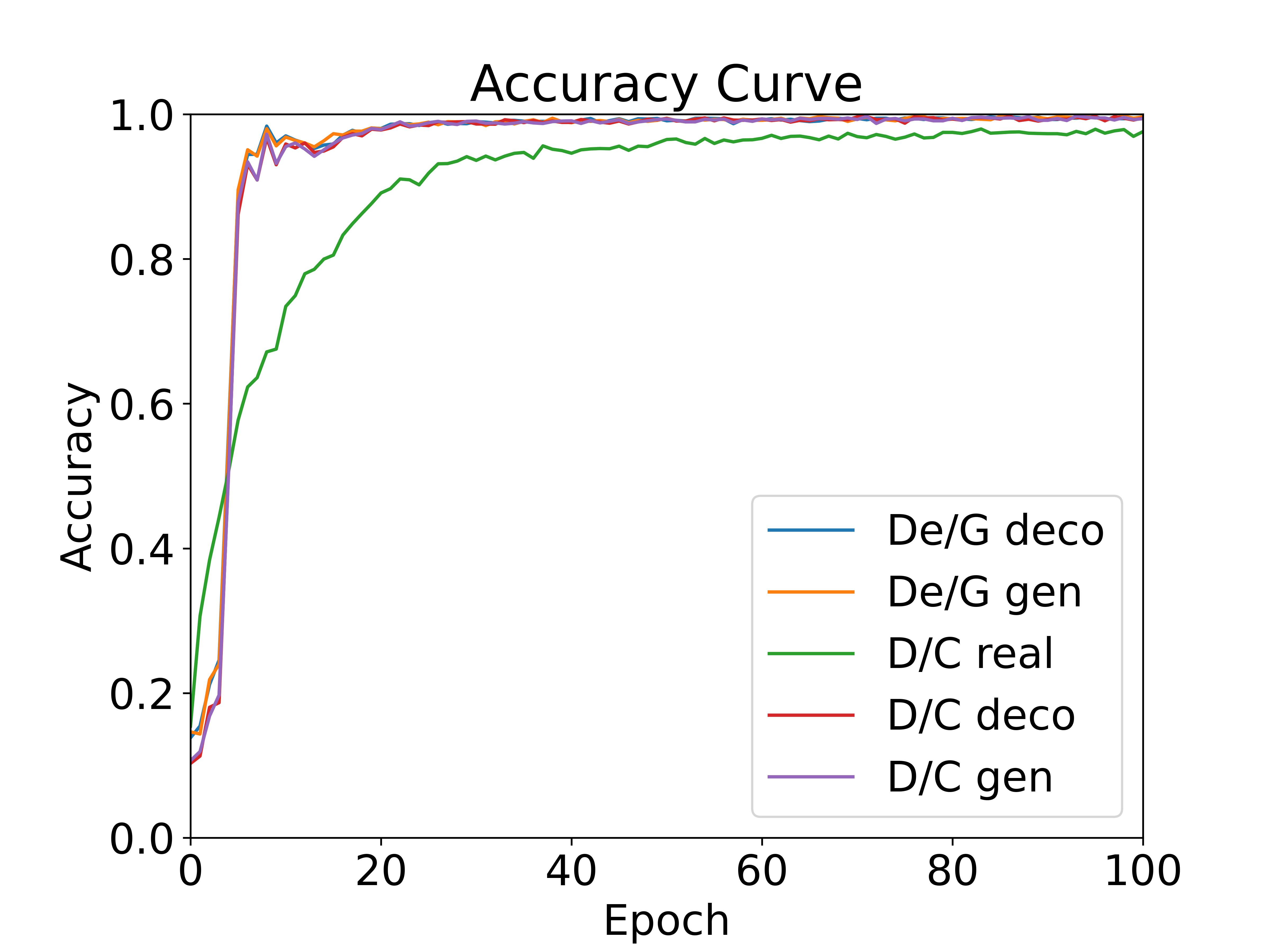}%
\label{fig:accuracy-2}}
\caption{\textcolor{black}{Loss curve and classification accuracy curve of AC-VAEGAN for the MSTAR dataset. (a) Loss curve. (b) Accuracy curve.}}
\label{fig:loss-accuracy-2}
\end{figure*}

\textcolor{black}{
Fig.~\ref{fig:loss-accuracy-2}\subref{fig:loss-2} and Fig.~\ref{fig:loss-accuracy-2}\subref{fig:accuracy-2} plot the loss curve and the classification accuracy curve of AC-VAEGAN in the first 100 training epochs for MSTAR dataset, respectively.
Again, from Fig.~\ref{fig:loss-accuracy-2}\subref{fig:loss-2} and Fig.~\ref{fig:loss-accuracy-2}\subref{fig:accuracy-2}, we can conclude that the training of AC-VAEGAN is stable and converges rapidly.
}

\subsection{Evaluation of the Synthetic Samples by AC-VAEGAN and AC-GAN}

\subsubsection{Evaluation of the Sea-Land Clutter Samples Synthesized by AC-VAEGAN and AC-GAN}

\begin{figure*}[!t]
\centering
\includegraphics[width=5in]{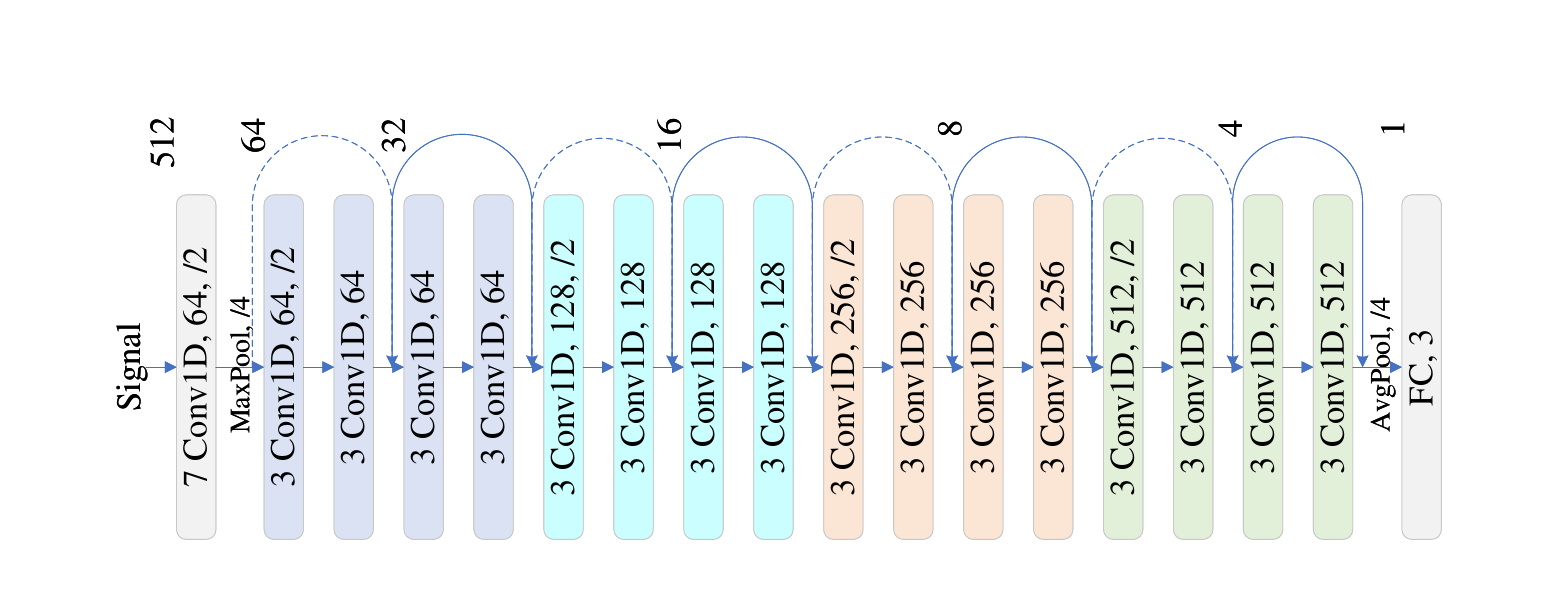}
\caption{Network structure of ResNet18 for sea-land clutter sample classification.}
\label{fig:ResNet18}
\end{figure*}

When using GAN-train and GAN-test to evaluate synthetic sea-land clutter samples, ResNet18~\cite{he2016deep} is employed.
The network structure and hyperparameter configurations of ResNet18 for sea-land clutter sample classification are shown in Fig.~\ref{fig:ResNet18} and Table~\ref{table:paramater classification}.
ResNet18 achieves 100\% training accuracy and 98.87\% testing accuracy on the original sea-land clutter dataset.

\begin{table}[!t]
\centering
\caption{The hyperparameter Configuration of ResNet18 for Sea-Land Clutter Sample Classification}
\label{table:paramater classification}
\begin{tabular}{cc}
\hline
Configuration & Default\\
\hline
Training Epoch & 100\\
Batch Size & 64\\
Learning Rate & 0.0001\\
Cross Entropy Loss & -\\
Adma Optimizer & beta1:0.9, beta2:0.999\\
Data Normalization & [-1,1]\\
Weight Initialization & -\\
\hline
\end{tabular}
\end{table}

\begin{figure*}
\centering
\subfloat[]{
\begin{minipage}[t]{0.3\textwidth}
\centering
\includegraphics[width=2in]{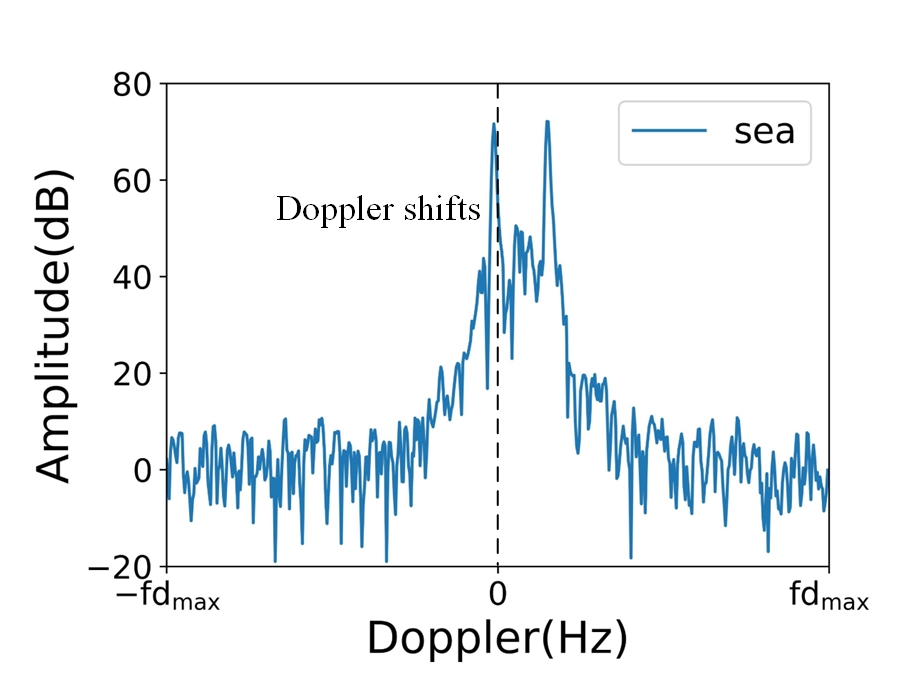}\\
\vspace{0.02cm}
\includegraphics[width=2in]{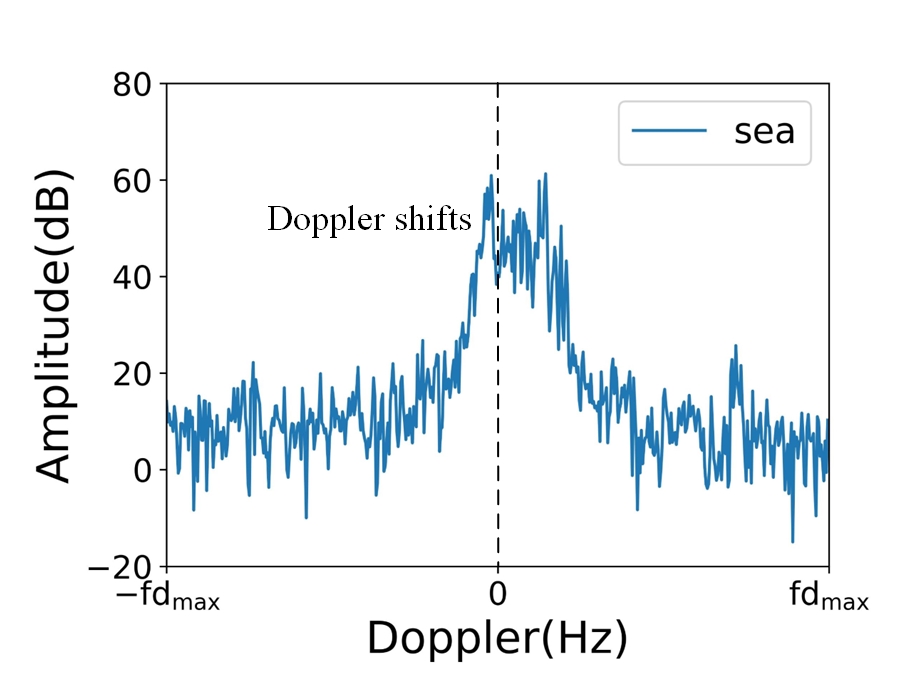}
\end{minipage}}
\subfloat[]{
\begin{minipage}[t]{0.3\textwidth}
\centering
\includegraphics[width=2in]{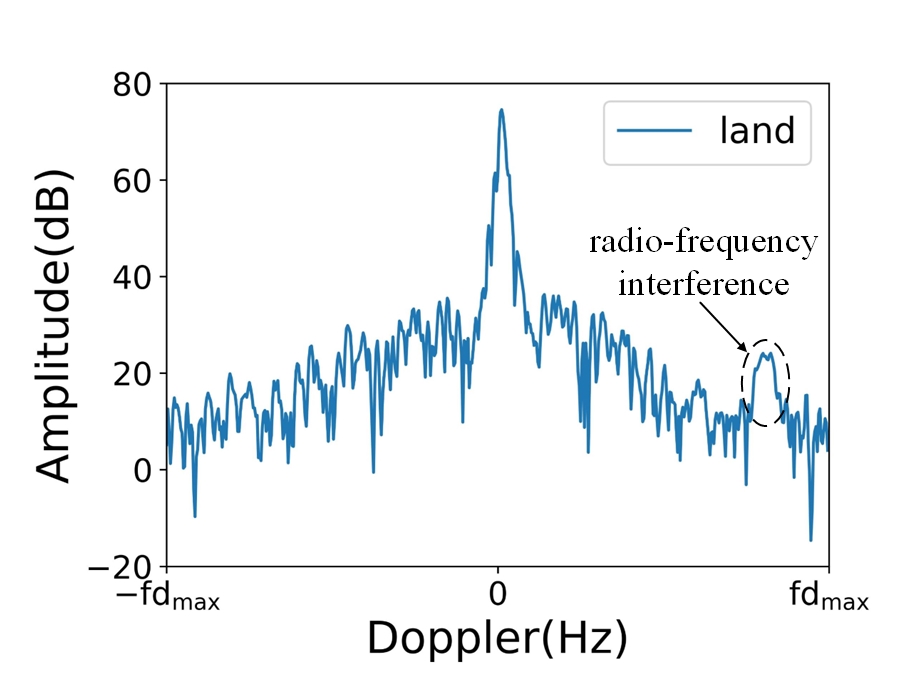}\\
\vspace{0.02cm}
\includegraphics[width=2in]{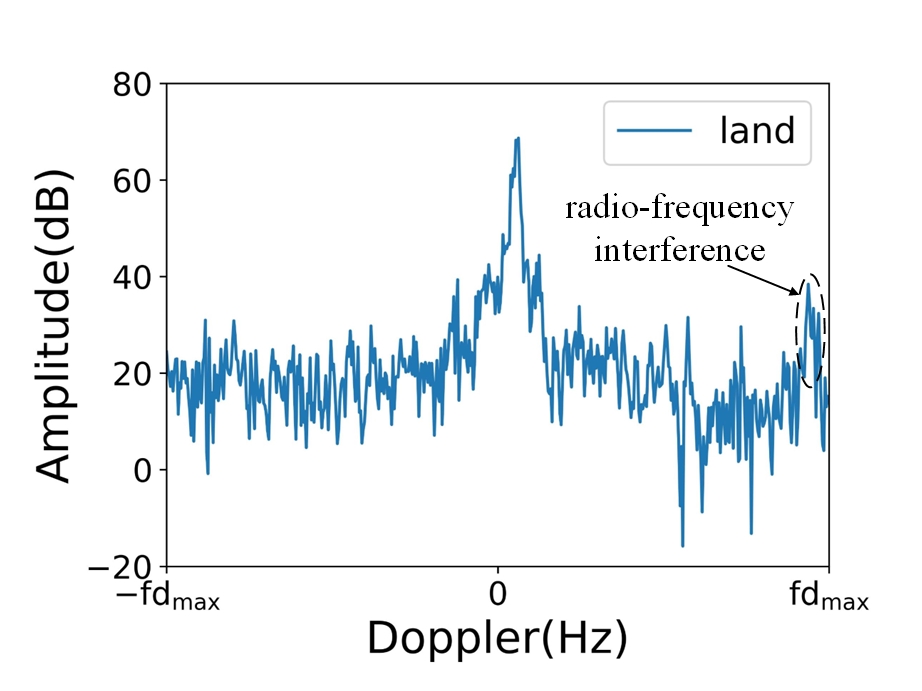}
\end{minipage}}
\subfloat[]{
\begin{minipage}[t]{0.3\textwidth}
\centering
\includegraphics[width=2in]{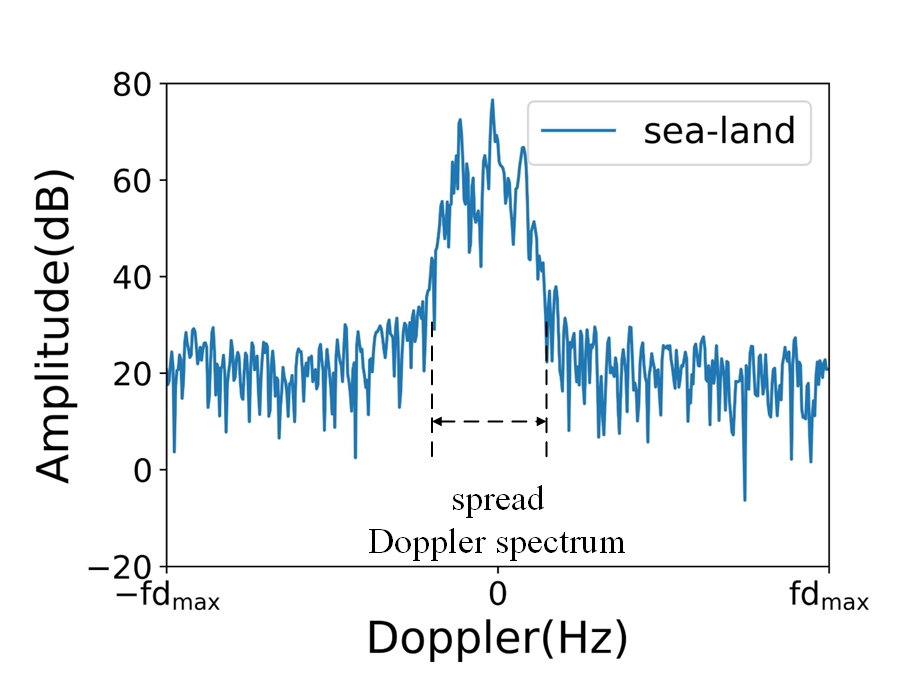}\\
\vspace{0.02cm}
\includegraphics[width=2in]{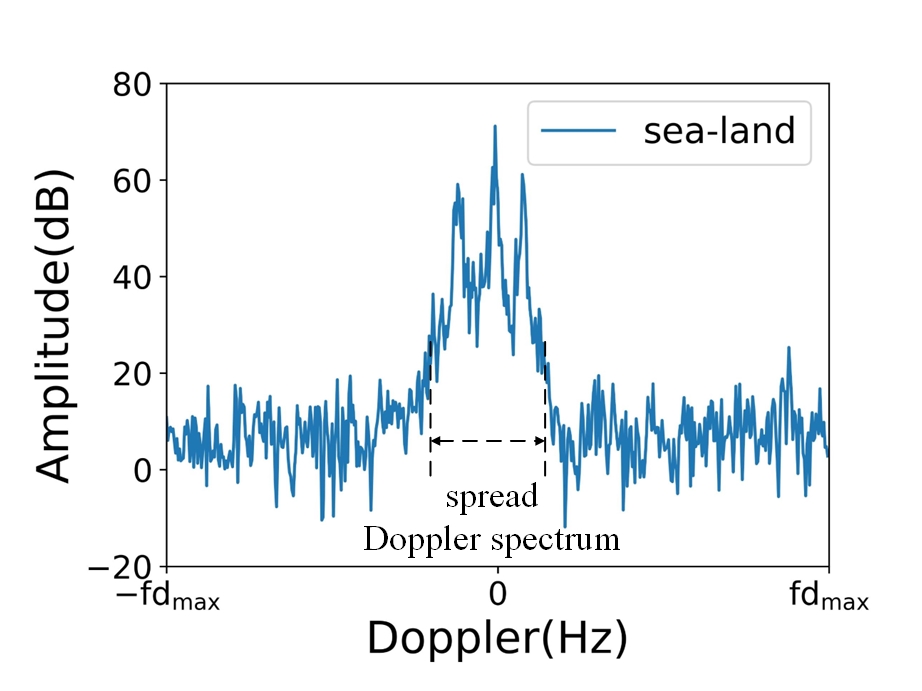}
\end{minipage}}

\caption{\textcolor{black}{The sea-land clutter samples synthesized by AC-VAEGAN: real samples~(top) and synthetic samples~(bottom). (a) sea clutter. (b) land clutter. (c) sea-land boundary clutter.}}
\label{fig:gen sample}
\end{figure*}

The trained AC-VAEGAN is used to synthesize 700 sea clutter samples, 700 land clutter samples, and 700 sea-land boundary clutter samples.
See Fig.~\ref{fig:gen sample} for the examples of fake samples.
Visually, most of the three types of synthetic samples have their own features, and are consistent with the real samples.
We mix the synthetic samples with real samples, and then ask two \textcolor{black}{human experts} to identify them.
The error rates of Expert 1 and Expert 2 misjudging synthetic samples as real samples are 90\% and 93\%, \textcolor{black}{real samples as synthetic samples are 10\% and 9\%}, respectively.

Further, the quality of the synthetic sea-land clutter samples is evaluated in terms of the metrics including GAN-train, GAN-test, AD, CS and PCC.
The evaluation results are shown in Table~\ref{table:synthesis evaluation}.
As far as GAN-train is concerned, we mix three types of synthetic sea-land clutter samples together as training data for ResNet18. Then, the original sea-land clutter test data is used as the test data for evaluation.
The average classification accuracy after training ResNet18 to a steady state is 76.44\%.
As far as GAN-test is concerned, we evaluate synthetic samples with ResNet18 trained on the original dataset and obtain the classification accuracy of 92.57\%.
This shows that the sea-land clutter samples synthesized by AC-VAEGAN have good diversity and fidelity.

From the perspective of statistical evaluation, the three types of synthetic sea-land clutter samples and the original training data are statistically analyzed, and the average evaluation results of AD, CS and PCC are obtained.
Different from GAN-train and GAN-test,
the three metrics evaluate the global properties of synthetic samples from a statistical signal perspective.
High scores are given to the samples synthesized by AC-VAEGAN for all the three statistical metrics.
The value of AD is 0.0160~dB.
The value of CS is 0.8988, which is very close to 1.
The value of PCC is 0.7897, showing that the signal has a strong correlation with real samples.
\textcolor{black}{
To further verify that the proposed statistical metrics are reasonable, we collect another sea-land clutter dataset according to the same way in Section \ref{subsec:Sea-Land Clutter Dataset} and evaluate the degree of similarity between the two real sea-land clutter datasets.
The results are as follows.
The value of AD is 0.0069~dB, the value of CS is 0.9428, and the value of PCC is 0.9183.}
These further verify that AC-VAEGAN is able to synthesize high quality sea-land clutter samples.

Besides, we compare the quality of sea-land clutter samples synthesized by AC-VAEGAN and AC-GAN.
AC-GAN removes En from AC-VAEGAN, and the rest of network structure and hyperparameter configurations are the same as AC-VAEGAN.
The evaluation results of AC-GAN are shown in Table \ref{table:synthesis evaluation}.
Comparing with the evaluation results of AC-VAEGAN and AC-GAN,
one can observe that the quality of sea-land clutter samples synthesized by AC-VAEGAN is better than that of AC-GAN.

\begin{table}[!t]
\centering
\caption{The Evaluation Results of Synthetic Sea-Land Clutter Samples}
\label{table:synthesis evaluation}
\begin{tabular}{c|ccc}
\hline
\hline
Evaluation Method & Metric & AC-GAN & AC-VAEGAN\\
\hline
\multirow{2}*{Traditional Evaluation} & GAN-train & 68.22\% & 76.44\% \\
~ & GAN-test & 81.29\% & 92.57\% \\
\hline
\multirow{3}*{Statistical Evaluation} & AD & 0.0161 & 0.0160 \\
~ & CS & 0.8036 & 0.8988 \\
~ & PCC & 0.7649 & 0.7897 \\
\hline
\hline
\end{tabular}
\end{table}

\subsubsection{\textcolor{black}{Evaluation of the MSTAR Samples Synthesized by AC-VAEGAN and AC-GAN}}
\begin{figure*}[!t]
\centering
\includegraphics[width=5in]{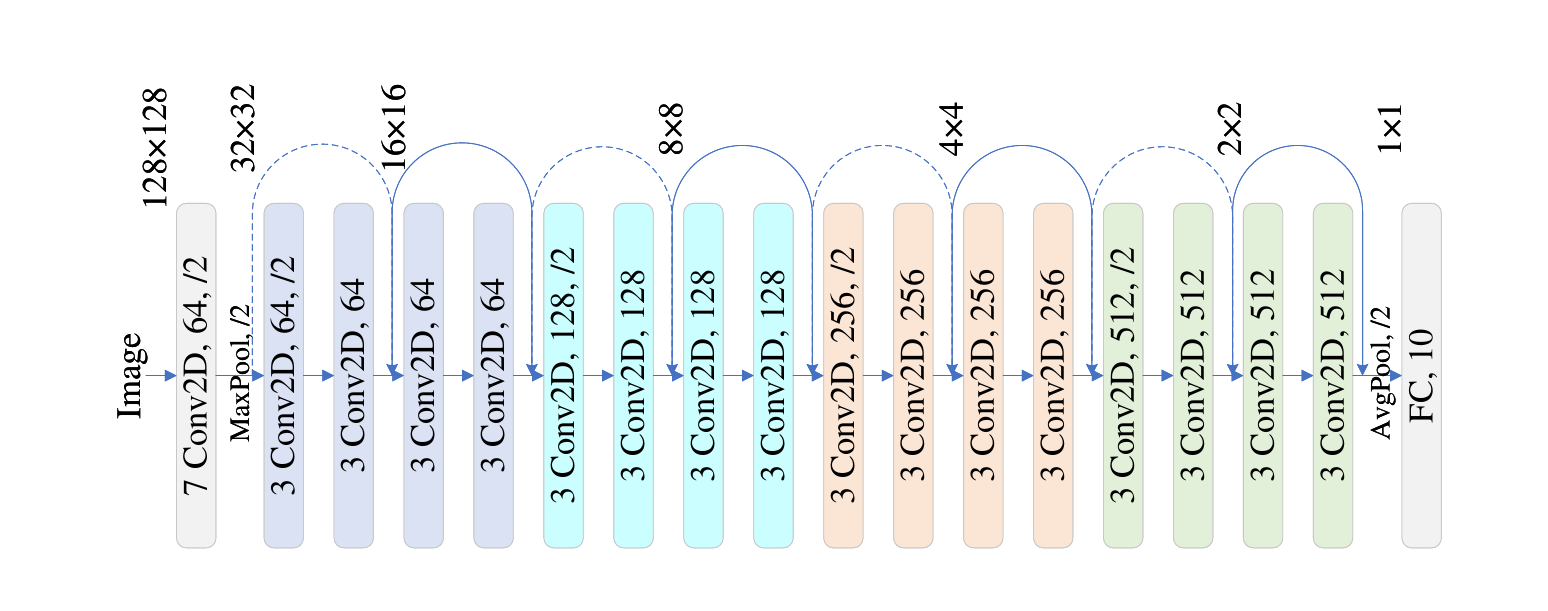}
\caption{\textcolor{black}{Network structure of ResNet18 for MSTAR sample classification.}}
\label{fig:ResNet18-2}
\end{figure*}

\textcolor{black}{
Since MSTAR dataset is an image dataset, we only use GAN-train and GAN-test to evaluate the quality of the synthetic MSTAR samples.
Firstly, the convolutional kernel of ResNet18 is adjusted to evaluate synthetic MSTAR samples~(see Fig.~\ref{fig:ResNet18-2}).
The hyperparameter configurations of ResNet18 for MSTAR dataset are the same as those of sea-land clutter dataset~(see Table~\ref{table:paramater classification}).
ResNet18 achieves 100\% training accuracy and 96.79\% testing accuracy on the original MSTAR dataset.}

\begin{figure*}
\centering
\subfloat[]{
\begin{minipage}[t]{0.15\textwidth}
\centering
\includegraphics[height=0.8in, width=0.8in]{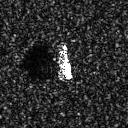}\\
\vspace{0.02cm}
\includegraphics[height=0.8in, width=0.8in]{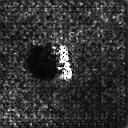}
\end{minipage}}
\subfloat[]{
\begin{minipage}[t]{0.15\textwidth}
\centering
\includegraphics[height=0.8in, width=0.8in]{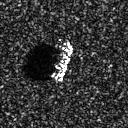}\\
\vspace{0.02cm}
\includegraphics[height=0.8in, width=0.8in]{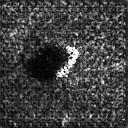}
\end{minipage}}
\subfloat[]{
\begin{minipage}[t]{0.15\textwidth}
\centering
\includegraphics[height=0.8in, width=0.8in]{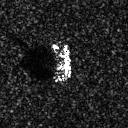}\\
\vspace{0.02cm}
\includegraphics[height=0.8in, width=0.8in]{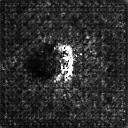}
\end{minipage}}
\subfloat[]{
\begin{minipage}[t]{0.15\textwidth}
\centering
\includegraphics[height=0.8in, width=0.8in]{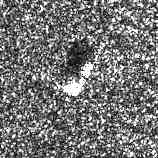}\\
\vspace{0.02cm}
\includegraphics[height=0.8in, width=0.8in]{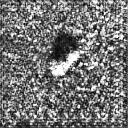}
\end{minipage}}
\subfloat[]{
\begin{minipage}[t]{0.15\textwidth}
\centering
\includegraphics[height=0.8in, width=0.8in]{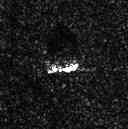}\\
\vspace{0.02cm}
\includegraphics[height=0.8in, width=0.8in]{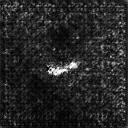}
\end{minipage}}

\vfil
\vspace{1pt}

\subfloat[]{
\begin{minipage}[t]{0.15\textwidth}
\centering
\includegraphics[height=0.8in, width=0.8in]{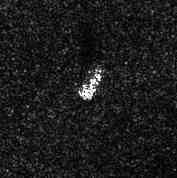}\\
\vspace{0.02cm}
\includegraphics[height=0.8in, width=0.8in]{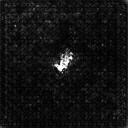}
\end{minipage}}
\subfloat[]{
\begin{minipage}[t]{0.15\textwidth}
\centering
\includegraphics[height=0.8in, width=0.8in]{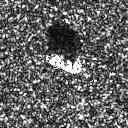}\\
\vspace{0.02cm}
\includegraphics[height=0.8in, width=0.8in]{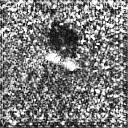}
\end{minipage}}
\subfloat[]{
\begin{minipage}[t]{0.15\textwidth}
\centering
\includegraphics[height=0.8in, width=0.8in]{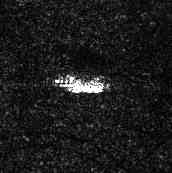}\\
\vspace{0.02cm}
\includegraphics[height=0.8in, width=0.8in]{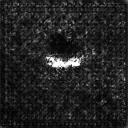}
\end{minipage}}
\subfloat[]{
\begin{minipage}[t]{0.15\textwidth}
\centering
\includegraphics[height=0.8in, width=0.8in]{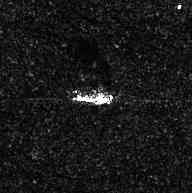}\\
\vspace{0.02cm}
\includegraphics[height=0.8in, width=0.8in]{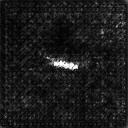}
\end{minipage}}
\subfloat[]{
\begin{minipage}[t]{0.15\textwidth}
\centering
\includegraphics[height=0.8in, width=0.8in]{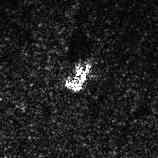}\\
\vspace{0.02cm}
\includegraphics[height=0.8in, width=0.8in]{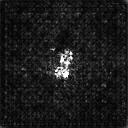}
\end{minipage}}

\caption{\textcolor{black}{
The MSTAR samples synthesized by AC-VAEGAN: real samples~(top) and synthetic samples~(bottom). (a) BMP2. (b) BTR70. (c) T72. (d) 2S1. (e) BRDM2. (f) D7. (g) BTR60. (h) T62. (i) ZIL131. (j) ZSU234.
}}
\label{fig:gen sample-2}
\end{figure*}

\textcolor{black}{
The trained AC-VAEGAN is used to synthesize 233 BMP2 samples, 233 BTR70 samples,
232 T72 samples, 299 2S1 samples, 298 BRDM2 samples, 299 D7 samples, 256 BTR60 samples, 299 T62 samples, 299 ZIL131 samples,
and 299 ZSU234 samples.
See Fig.~\ref{fig:gen sample-2} for the examples of fake samples.
Again, we remove En and use AC-GAN to synthesize the same number of MSTAR samples as above.
Then, the quality of MSTAR samples synthesized by AC-VAEGAN and AC-GAN are compared.
The evaluation results are shown in Table \ref{table:synthesis evaluation-2}.
Comparing with the evaluation results of AC-VAEGAN and AC-GAN,
one can observe that the quality of MSTAR samples synthesized by AC-VAEGAN is better than that of AC-GAN.
}

\begin{table}[!t]
\centering
\caption{\textcolor{black}{The Evaluation Results of Synthetic MSTAR Samples}}
\label{table:synthesis evaluation-2}
\begin{tabular}{c|ccc}
\hline
\hline
Evaluation Method & Metric & AC-GAN & AC-VAEGAN\\
\hline
\multirow{2}*{Traditional Evaluation} & GAN-train & 60.74\% & 63.07\% \\
~ & GAN-test & 66.40\% & 69.28\% \\
\hline
\hline
\end{tabular}
\end{table}

\subsection{\textcolor{black}{Data Augmentation Experiment Based on AC-VAEGAN and AC-GAN}}
\subsubsection{\textcolor{black}{Data Augmentation Experiment of sea-land clutter dataset Based on AC-VAEGAN and AC-GAN}}
\textcolor{black}{
Finally, the impact of data augmentation based on \textcolor{black}{AC-VAEGAN/AC-GAN} on the performance improvement of ResNet18 is evaluated in the cases of imbalanced and scarce sea-land clutter samples.
We conduct six groups of comparative experiments.
The first three groups are imbalanced sample experiments, which are based on the imbalanced datasets of NO.1-NO.3 in Table \ref{table:imbalance and scarce samples}.
The last three groups are scarce sample experiments, which are based on the scarce datasets of NO.4-NO.6 in Table~\ref{table:imbalance and scarce samples}.
The steps of the experiments are as follows:
(1) Select the data augmentation dataset NO.k, k$\in\{1,2,...6\}$; (2) Train \textcolor{black}{AC-VAEGAN/AC-GAN} based on the current dataset NO.k;
(3) Use the trained \textcolor{black}{AC-VAEGAN/AC-GAN} to perform data augmentation on minority class samples to obtain a balanced sea-land clutter dataset, denoted as \textcolor{black}{NO.k$^{++}$/NO.k$^{+}$};
(4) Treat NO.k and \textcolor{black}{NO.k$^{++}$/NO.k$^{+}$} as training data, and original sea-land clutter test data as test data. Then, the average classification accuracy after training ResNet18 to a steady state is recorded, and the best one after 100 repetitions of training is taken as the final test result.}

\begin{figure}[!t]
\centering
\includegraphics[width=3.5in]{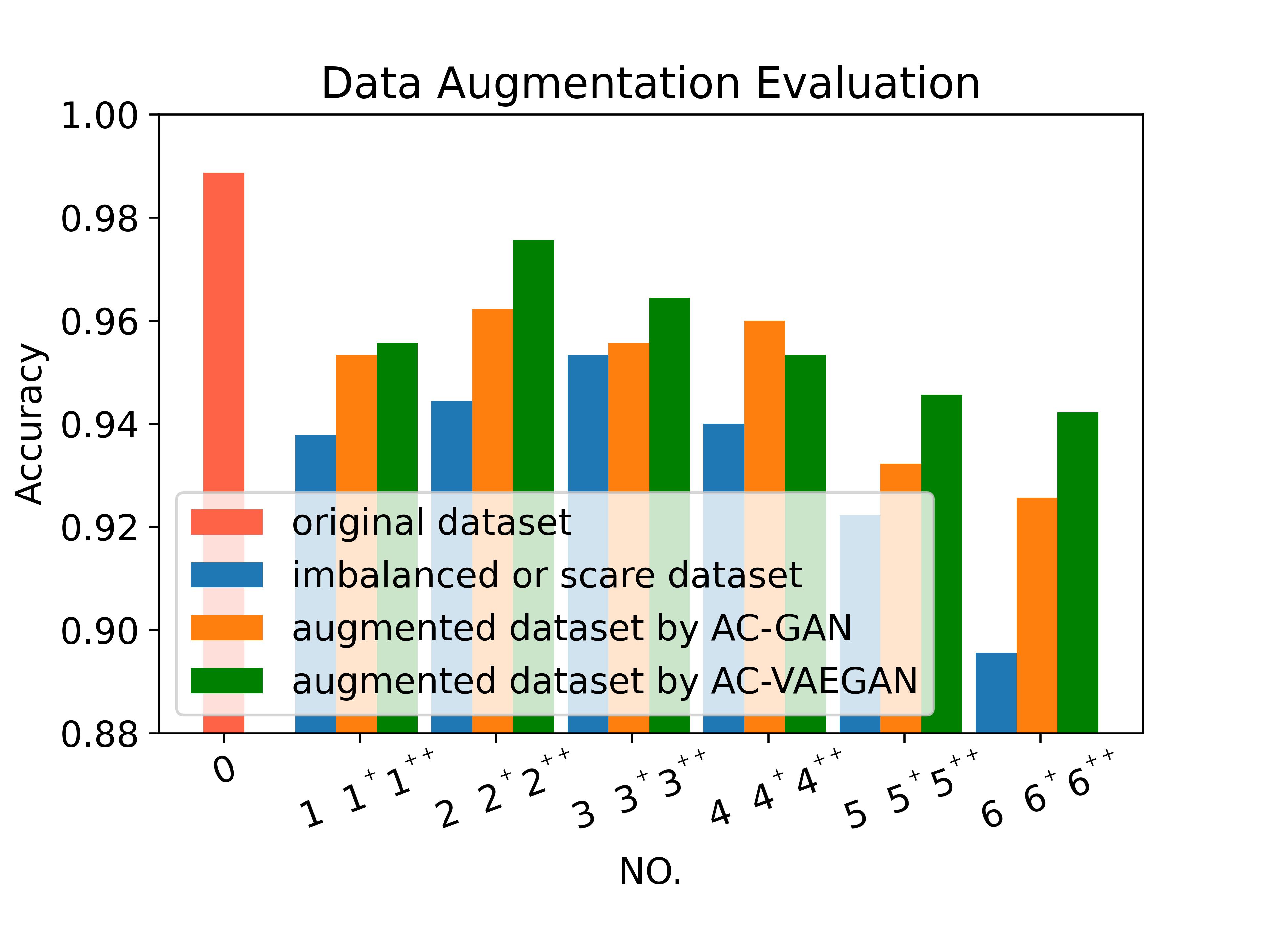}
\caption{\textcolor{black}{Data augmentation evaluation results of the sea-land clutter dataset based on AC-VAEGAN and AC-GAN.}}
\label{fig:data augmentation}
\end{figure}

\textcolor{black}{
Fig.~\ref{fig:data augmentation} plots the classification accuracy of NO.0-NO.6, NO.1$^{+}$-NO.6$^{+}$ and NO.1$^{++}$-NO.6$^{++}$ tested on ResNet18, where NO.0 represents the classification accuracy of the original dataset~(98.87\%).
Comparing with the classification accuracy of NO.0, the classification performance of NO.1-NO.6 is reduced.
}
It is seen from the comparison results that the classification accuracy of sea-land clutter samples after data augmentation has been improved.
NO.1-NO.3 represent the imbalanced dataset of sea-land boundary, land, and sea clutter samples as the minority, respectively.
It is seen that AC-VAEGAN data augmentation method can improve the classification performance in the case of imbalanced samples. NO.4-NO.6 represent sea-land clutter datasets from light to heavy scarcity, respectively.
It is seen that AC-VAEGAN data augmentation method can improve the classification performance in the case of scarce samples, and as the sample scarcity increases, the improvement on the classification accuracy is more obvious.
\textcolor{black}{Besides, the data augmentation performance of AC-VAEGAN is generally superior to that of AC-GAN.}
In conclusion, the proposed AC-VAEGAN is able to serve as an effective tool for data augmentation in sea-land clutter classification.

\textcolor{black}{
In this paper, we focus on the classification of land/sea clutter.
The purpose is to match the derived classification results with a prior geographic information, and then provide coordinate registration parameters for target localization.
Since, in a typical OTHR,
the land/sea clutter is much stronger
(typically 30–50 dB) than the signal of targets,
we treat the targets' signal as noise or interference in the classification of land/sea clutter, no matter how close the targets' Doppler to the clutter.
In our experiments,
we do not differentiate the clutter samples with or without targets' signal.
Note that target detection is another important issue in OTHR.
We leave the extension of our proposed method to
target detection while target's Doppler are close to the sea/land clutter as future work.
}

\subsubsection{\textcolor{black}{Data Augmentation Experiment of MSTAR Dataset Based on AC-VAEGAN and AC-GAN}}
\begin{figure}[!t]
\centering
\includegraphics[width=3.5in]{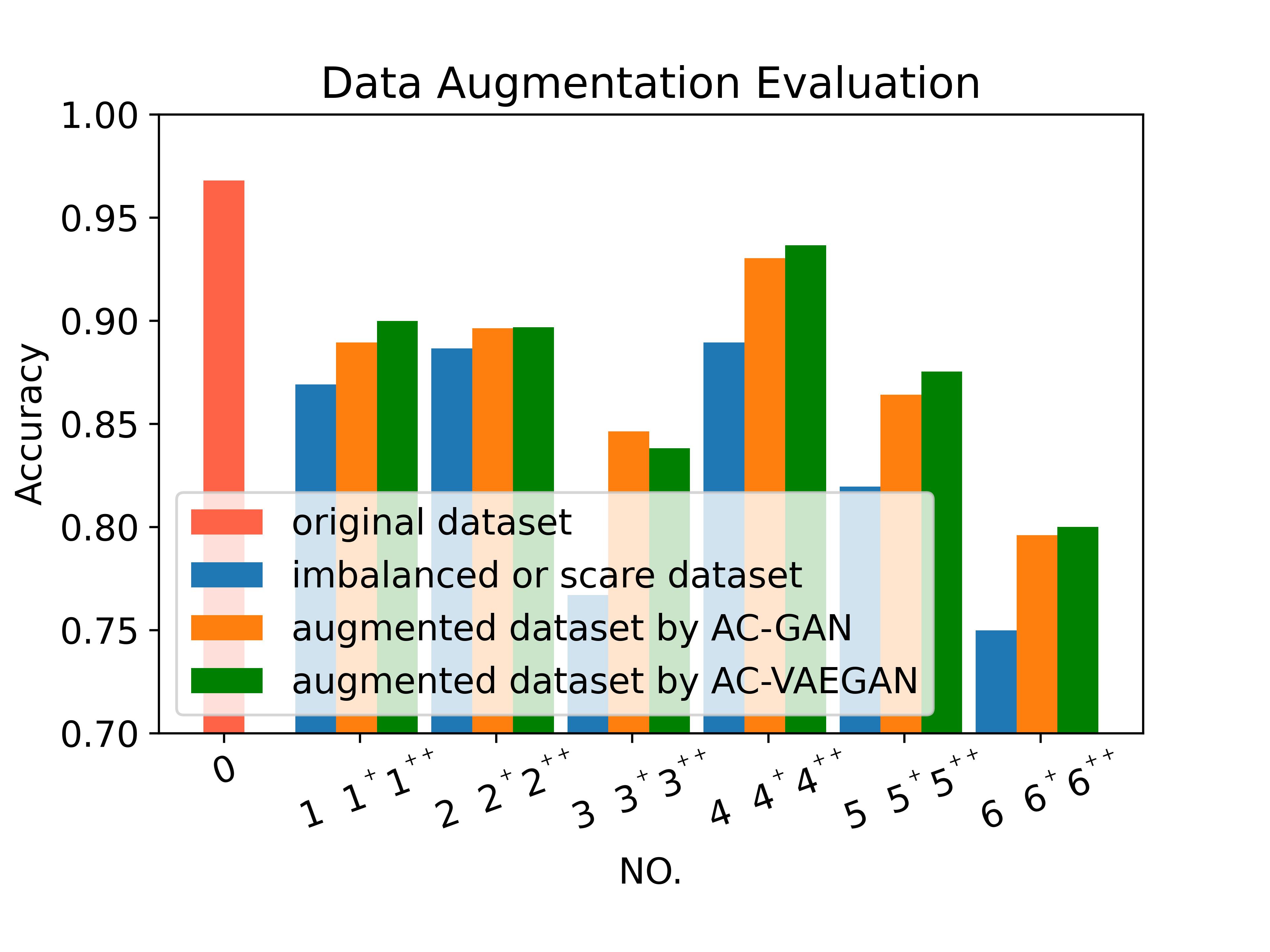}
\caption{\textcolor{black}{Data augmentation evaluation results of the MSTAR dataset based on AC-VAEGAN and AC-GAN.}}
\label{fig:data augmentation-2}
\end{figure}

\textcolor{black}{
Similarly, following the steps described above for data augmentation of sea-land clutter, the impact of data augmentation based on AC-VAEGAN/AC-GAN on the performance improvement of ResNet18 is evaluated in the cases of imbalanced and scarce MSTAR samples in Table \ref{table:imbalance and scarce samples-2}.
The data augmentation evaluation results are shown in Fig.~\ref{fig:data augmentation-2}, which further verifies the data augmentation performance of the proposed AC-VAEGAN.}

\section{Conclusions and Future Work}
\label{sec:Conclusions and Future work}
A novel network, namely AC-VAEGAN, was proposed to act as a data augmentation method for sea-land clutter classification of OTHR.
An evaluation method combining traditional evaluation of GAN domain and statistical evaluation of signal domain was proposed to evaluate the quality of synthetic samples.
Experimental results demonstrated that AC-VAEGAN can synthesis higher quality samples than AC-GAN.
Using the samples synthesized by AC-VAEGAN as the supplement of \textcolor{black}{imbalanced and scarce datasets} can improve the performance of sea-land clutter classification model.

In future work, there are two issues that need to be addressed:
(1) The loss function of AC-VAEGAN does not contain an interpretable indicator to guide the training process like
that of classifier. Therefore, it is necessary to improve the loss
function of AC-VAEGAN by combining Wasserstein distance
and other loss functions;
(2) GAN also has high requirements on the quality of training data. In fact, there are more complex sea/land clutter samples from OTHR. Therefore, we intend to study how to reduce the high quality requirements of GAN for training data.
\textcolor{black}{
Other potential areas for future research include the classification of more complicated sea/land clutter and extension of the proposed network to target detection.
}

%\section*{Acknowledgments}

\bibliographystyle{IEEEtran}
\bibliography{references}

\end{document}